\documentclass{aa}  
\usepackage{graphicx}
\usepackage{caption}
\usepackage[varg]{txfonts}
\usepackage[colorlinks=true,linkcolor=blue,citecolor=blue,urlcolor=blue]{hyperref}
\makeatletter
\renewcommand*\aa@pageof{, page \thepage{} of \pageref*{LastPage}}
\makeatother
\usepackage{color}
\usepackage{nicefrac}
\usepackage{textcomp}

\bibpunct{(}{)}{;}{a}{}{,} 
\usepackage{comment}
\usepackage{siunitx}
\usepackage{mathrsfs}
\usepackage{placeins}
\usepackage{float}

\begin{document}

\title{Correlation between the optical veiling and accretion properties}

\subtitle{A case study of the classical T\,Tauri star DK\,Tau}

\author{M.\,Nelissen\inst{1} \and A.\,Natta\inst{1} \and P.\,McGinnis\inst{1} \and C.\,Pittman\inst{2} \and C.\,Delvaux\inst{3} \and T.\,Ray\inst{1}}

\institute{Dublin Institute for Advanced Studies, Astronomy \& Astrophysics Section, 
31 Fitzwilliam Place, Dublin\,2, Ireland \\
\email{nelissen@cp.dias.ie} 
\and Institute for Astrophysical Research, Department of Astronomy, Boston University, 725 Commonwealth Avenue, Boston, MA 02215, USA
\and Max-Planck-Institut für extraterrestrische Physik, Giessenbachstr. 1, 85748 Garching, Germany}

\date{Accepted 25 July 2023}

\abstract{
    {\textit{Context.}  Classical T\,Tauri stars (cTTs) accrete from their circumstellar disk. The material falls onto the stellar surface, producing an accretion shock, which generates veiling in a star's spectra. In addition, the shock causes a localized accretion spot at the level of the chromosphere. 
    }
    
    {\textit{Aims.} Our goal is to investigate the accretion, particularly the mass accretion rates ($\dot{M}_{\text{acc}}$), for the cTTs DK\,Tau, over two periods of 17 and 29 days, using two different procedures for comparison purposes. 
    }
    
    {\textit{Methods.} The first method relies on the derivation of the accretion luminosity via accretion-powered emission lines. The second compares the variability of the optical veiling with accretion shock models to determine mass accretion rates. 
    We used observations taken in 2010 and 2012 with the ESPaDOnS (Echelle SpectroPolarimetric Device for the Observation of Stars) spectropolarimeter at the CFHT (Canada-France-Hawaii Telescope). 
   }
    
    {\textit{Results.} We find peak values of the veiling (at $\sim$550\,nm) ranging from 0.2 to 1.3, with a steeper trend across the wavelength range for higher peak values. When using the accretion-powered emission lines, we find mass accretion rate values ranging from log\,($\dot{M}_{\text{acc}}$[$M_\odot$\,yr$^{-1}$]) = -8.20 to log\,($\dot{M}_{\text{acc}}$[$M_\odot$\,yr$^{-1}$]) = -7.40. This agrees with the values found in the literature, as well as the values calculated using the accretion shock models and the veiling. 
    In addition, we identify a power-law correlation between the values of the accretion luminosity and the optical veiling. 
    For the 2010 observations, using the values of the filling factors (which represent the area of the star covered by an accretion spot) derived from the shock models, we infer that the accretion spot was located between +45\si{\degree} and +75\si{\degree} in latitude. 
    }
    
    {\textit{Conclusion.} We show that both methods of determining the mass accretion rate yield similar results. We also present a helpful means of confirming the accretion luminosity values by measuring the veiling at a single wavelength in the optical. 
    }
}

\keywords{Stars: individual: DK\,Tau - Stars: low-mass - Stars: variables: T Tauri - Stars: formation - Accretion, accretion disks - Techniques: spectroscopic}

\maketitle


\section{Introduction} \label{sec:introduction}

Classical T\,Tauri stars (cTTs) are low-mass ($M_\star$ $\leqslant$ 2\,$M_\odot$), optically visible, pre-main sequence stars that are magnetically active and are accreting from their surrounding disks. These circumstellar disks are the birth place of exoplanets, with the era of planet formation overlapping that of star formation. Accretion shapes the lifetime of the disks, which is on the order of a few million years \cite[see][]{2016ARA&A..54..135H}. Accretion is also thought to have a significant impact on the development of young stars by transferring mass, by impacting angular momentum evolution, and by being involved in jet launching \cite[see][]{2014prpl.conf..451F}. Understanding the accretion process is thus an important issue for the establishment of plausible stellar and planet formation theories. cTTs present an opportunity to probe the interactions between the stars, their magnetic fields, and their disks through the accretion of circumstellar material. 

According to the current magnetospheric accretion paradigm \cite[see e.g.,][]{1994ApJ...429..781S, 2002ApJ...578..420R, 2008A&A...478..155B, 2016ARA&A..54..135H}, the magnetic field of a cTTs truncates the inner circumstellar disk at a few stellar radii and forces the migrating circumstellar material into accretion columns. The matter subsequently falls onto the stellar surface at near free-fall velocities, producing an accretion shock. These shocks generate emission lines and an excess continuum that is superimposed on the photospheric spectrum, which causes veiling of the photospheric absorption lines \cite[see][]{1998ApJ...509..802C}. This happens mostly at ultraviolet (UV) and optical wavelengths, with the intensity of the veiling decreasing at longer wavelengths, and this causes the absorption lines to appear shallower. High-resolution spectroscopy allows us to resolve the absorption lines and measure their decrease in depth due to veiling. The values of veiling experience temporal variations as well: as the star rotates, the accretion shock is seen at different angles with respect to the line of sight of the observer. With these ideas in mind, veiling can thus be used as a tracer of accretion activity and variability. 

The shocks also cause the emergence of localized accretion spots (also called bright or hot spots) at the chromospheric level, which are the footprints of the accretion columns on the surface of the star. It has been found that these accretion spots present a radial density gradient: a denser core surrounded by a low-density region \cite[see][]{2021Natur.597...41E}. However, details of the accretion process are still poorly understood \cite[see e.g.,][]{2019ApJ...874..129R, 2021Natur.597...41E, 2022AJ....163..114E, 2022A&A...667A.124G, 2022AJ....164..201P}. 

The mass accretion rate onto a star ($\dot{M}_{\text{acc}}$) is an important parameter when studying the evolution of the pre-main sequence stars and their disks. One method of determining this quantity is by using the accretion luminosity ($L_{\text{acc}}$), knowing the stellar mass and radius \cite[see][]{1998ApJ...492..323G}. By extrapolation, $L_{\text{acc}}$ can be derived from several accretion-powered emission lines, using empirical relations between the accretion luminosity and the line luminosities \cite[see][]{2017A&A...600A..20A}. 

A second method of extracting $\dot{M}_{\text{acc}}$ is by comparing the values of veiling to accretion shock models. These models were first introduced by \cite{1998ApJ...509..802C}. They modeled the base of the accretion column at the stellar surface using a geometry that is one-dimensional and plane-parallel. The accreting material is assumed to be at the free-fall velocity, which depends on the stellar mass and radius. Historically, the models were characterized by a single pair of energy flux and filling factor values, with the energy flux being the flux of energy carried by the accretion column and the filling factor being the fraction of the surface of the star covered by the accretion spot. \cite{2013ApJ...767..112I} updated the models by using multiple energy fluxes, each one with its corresponding filling factor, allowing for multiple accretion columns. Further improvements of the models have been made more recently \cite[see e.g.,][]{2019ApJ...874..129R, 2022AJ....163..114E, 2022AJ....164..201P}. 

In addition to its use with accretion shock models, veiling in the optical band can also be used to infer accretion luminosity, as \cite{2022A&A...668A..94S} have found in the case of the cTTs RU\,Lup. Furthermore, the evolution of various accretion-related quantities with stellar rotation can offer more insight into the accretion process. In this context, this work is a study of the accretion properties and variability in a particular star, DK\,Tau. 

DK\,Tau is a low-mass cTTs located in the Taurus molecular cloud at a distance of 132.6\,pc \cite[see][]{gaia1, gaia2, 2022arXiv220800211G} and accreting from its circumstellar disk. It shows considerable veiling \cite[see e.g.,][]{hartigan95, 2011ApJ...730...73F}. It is in a wide binary system \cite[i.e., separated by 2.38\arcsec - see e.g.,][]{2019A&A...628A..95M}. This separation allows DK\,Tau A (called "DK\,Tau" for short hereafter) to be spatially resolved by the ESPaDOnS (Echelle SpectroPolarimetric Device for the Observation of Stars) spectropolarimeter and investigated on its own. DK\,Tau has a K7 spectral type \cite[see e.g.,][]{2007ApJ...664..975J, 2011ApJ...730...73F}, a 4\,000\,K effective temperature, an 8.2\,days rotation period, a 2.48\,$R_{\odot}$ radius, and its rotation axis is inclined by 58\si{\degree} \cite[see][]{2023A&A...670A.165N}. Its outer disk is inclined by 21\si{\degree} \cite[see][]{2022arXiv220103588R}, and is misaligned compared to the inner disk and rotation axis \cite[see][]{2023A&A...670A.165N}. \cite{2007ApJ...664..975J} derived a mass of 0.7\,$M_\odot$ for DK\,Tau. In addition to accretion, the star also shows evidence of ejection \cite[see e.g.,][]{hartigan95}, in particular inner disk winds and a jet, which was revealed by the detection of both low and high velocity forbidden [OI] emission by \cite{2019ApJ...870...76B}. \cite{2007A&A...461..183G} reported a long-term photometric variability of $\overline{\Delta V}$ = 1.86\,mag. 

We present a study of the active accretor DK\,Tau, specifically of the variability of its veiling from night to night and across the visible wavelength range. We describe our observations and data analysis in Sect.\,\ref{sec:observations}. We detail our results regarding the accretion luminosity, the use of accretion shock models, and the calculations of the mass accretion rate in Sect.\,\ref{sec:results}. In Sect.\,\ref{sec:discussion}, we discuss a correlation between the veiling and the accretion properties, as well as variability in relation to stellar rotation and accretion spots. Finally, we present our main conclusions in Sect.\,\ref{sec:conclusions}. 


\section{Observations and data analysis} \label{sec:observations}

\subsection{ESPaDOnS Data} \label{Data}

We used observations of DK\,Tau taken with ESPaDOnS, an echelle spectropolarimeter covering the optical domain, from 370 to 1\,050\,nm, in a single exposure. ESPaDOnS is mounted at the CFHT (Canada-France-Hawaii Telescope), a 3.6 meter telescope in Hawaii. It has a fiber aperture of 1.66\arcsec and has a resolving power of 65\,000 \citep[see][]{2006ASPC..358..362D}. 

Two sets of nine circularly polarized spectra were collected in December 2010, and from the end of November to the end of December 2012. In 2010 the observations were taken over 17 days, and in 2012 the observations were taken over 29 days. This was done in order to capture at least one full rotation of DK\,Tau for each set. These observations were obtained as part of the MaPP (Magnetic Protostars and Planets) large program at the CFHT, under proposals 10BP12 and 12BP12, with J.-F.\,Donati as P.I. and are public. We downloaded the observations from the archive of the PolarBase website\footnote{\url{http://polarbase.irap.omp.eu}} \cite[see e.g.,][]{2014PASP..126..469P}, as well as the corresponding image files from the Canadian Astronomy Data Centre (CADC) website\footnote{\url{http://www.cadc-ccda.hia-iha.nrc-cnrc.gc.ca/en}}.  

The data had been previously reduced at the CFHT with the LibreESpRIT (for "Echelle Spectra Reduction: an Interactive Tool") reduction package specifically built for extracting polarization echelle spectra from raw data. This involved subtracting the bias and the dark frames, and correcting for the variations in sensitivity using flat field frames \citep[see][]{1997MNRAS.291..658D}. In addition to the LibreESpRIT automatic continuum normalization, we continuum normalized the spectra. The automatic procedure did not properly adjust the continuum, as it is not tailored for stars that display emission lines. 

The Julian date corresponding to the middle of each exposure  are listed in Table\,\ref{TableDates}. The total exposure time is 4\,996.0\,s for each observation. It is important to note that on 19 December 2010 the full Moon was close to DK\,Tau \cite[see][]{2023A&A...670A.165N} and, because of the resultant scattered light, we ignore this night in our analysis. 

\begin{table}
\begin{center}
\caption{Dates for the 2010 and 2012 datasets. \label{TableDates}}
\begin{tabular}{c c c}
\hline
\hline 
Date & Julian date & Rotation phase\\
(yyyy-mm-dd) &     & (8.2 day period) \\
\hline 
2010-12-14 & 2\,455\,544.979\,74 & 0.00 \\       
2010-12-15 & 2\,455\,545.853\,82 & 0.11 \\     
2010-12-16 & 2\,455\,546.854\,61 & 0.23 \\
2010-12-17 & 2\,455\,547.826\,04 & 0.35 \\
2010-12-18 & 2\,455\,548.819\,03 & 0.47 \\
2010-12-19 & 2\,455\,549.786\,41 & 0.59 \\
2010-12-24 & 2\,455\,554.883\,63 & 1.21 \\
2010-12-26 & 2\,455\,557.034\,94 & 1.47 \\
2010-12-30 & 2\,455\,560.977\,48 & 1.95 \\
\hline
2012-11-25 & 2\,456\,256.919\,80 & 0.00 \\       
2012-11-28 & 2\,456\,259.895\,31 & 0.36 \\     
2012-11-29 & 2\,456\,260.991\,38 & 0.50 \\
2012-12-01 & 2\,456\,262.947\,48 & 0.74 \\
2012-12-02 & 2\,456\,263.865\,70 & 0.85 \\
2012-12-04 & 2\,456\,265.965\,15 & 1.10 \\
2012-12-07 & 2\,456\,268.846\,50 & 1.45 \\
2012-12-10 & 2\,456\,271.825\,04 & 1.81 \\
2012-12-23 & 2\,456\,284.762\,49 & 3.40 \\
\hline 
\end{tabular}
\end{center}
\end{table}


\subsection{Optical veiling} \label{Veiling}

Because accretion shocks are at a higher temperature than the photosphere, they add extra flux to the stellar continuum. This artificially decreases the depth of photospheric absorption lines, an effect known as veiling. The veiling ($R$) is defined as the ratio between the accretion shock flux and the photospheric flux. The values of veiling vary as a function of wavelength for a single night, and also vary from night to night. 

In order to determine the values of the veiling in our spectra, we adopted a technique based on the fitting of a rotationally broadened and artificially veiled weak-lined T\,Tauri star (wTTs) template to the spectrum of DK\,Tau. We used an ESPaDOnS spectrum because of its high resolution and choose a wTTs with the same spectral type as our cTTs and coming from the same star forming region. This allows us to assume that both T\,Tauri stars (TTs) would have the same chemical composition, very similar age and $\log g$, and that the microturbulence and macroturbulence velocities should be very similar. Consequently, this wTTs can be seen as the purely photospheric version of DK\,Tau because it experiences no accretion. In addition, the wTTs also needs to have a $v \sin i$, the line of sight projected equatorial rotational velocity, lower than the $v \sin i$ of DK\,Tau, which is of 13.0\,km\,s$^{-1}$ \cite[see][]{2023A&A...670A.165N}. Indeed, the wTTs spectrum will be artificially broadened to match DK\,Tau in the fitting process and this cannot be done if the wTTs lines are already broader because of rotational broadening. The rotational broadening is performed by convolving the spectrum by a Doppler rotational profile as described in \cite{2008oasp.book.....G}. After exploring several ESPaDOnS spectra of wTTs with a K7 spectral type in order to find the one that provided the best fit, we ultimately used TAP45 \cite[which has a $v \sin i$ of 11.5\,km\,s$^{-1}$ - see][]{1987AJ.....94.1251F, 1993AAS..101..485B}. 

\begin{figure*}[hbtp]
\centering
\includegraphics[scale=0.39]{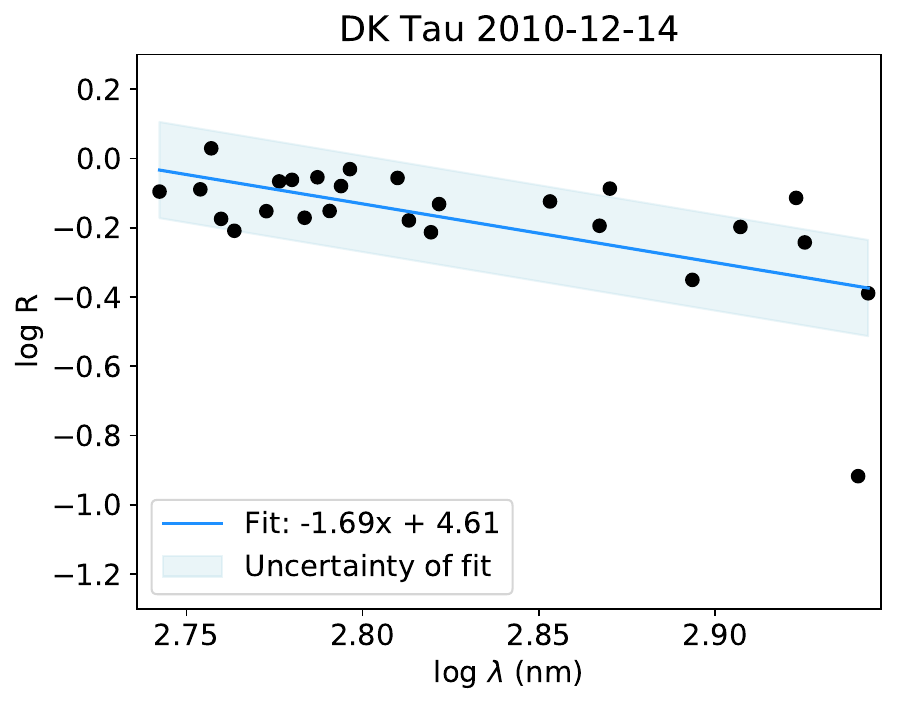}
\includegraphics[scale=0.39]{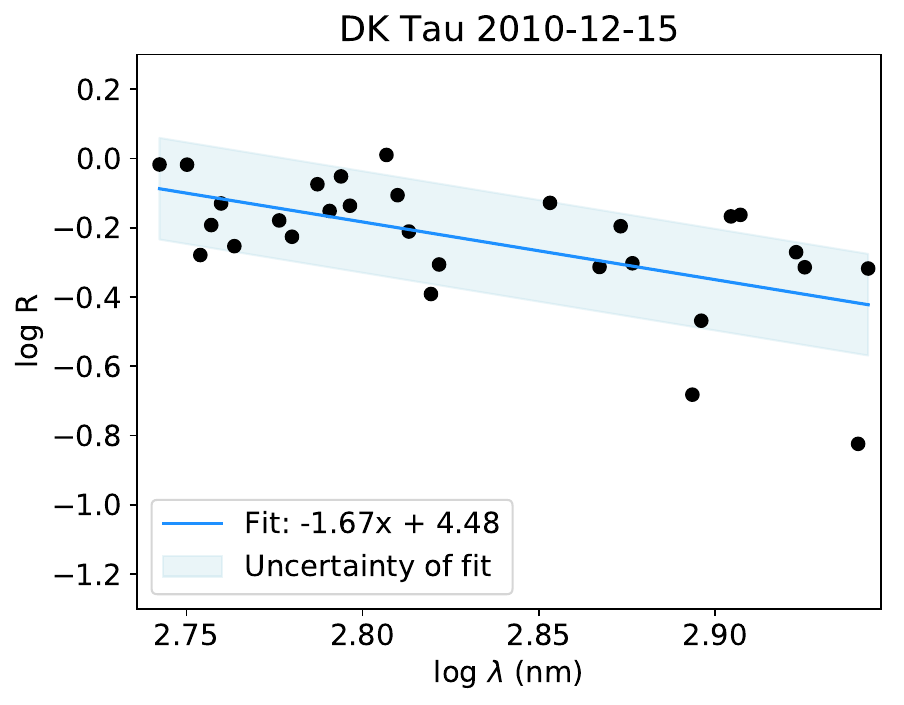}
\includegraphics[scale=0.39]{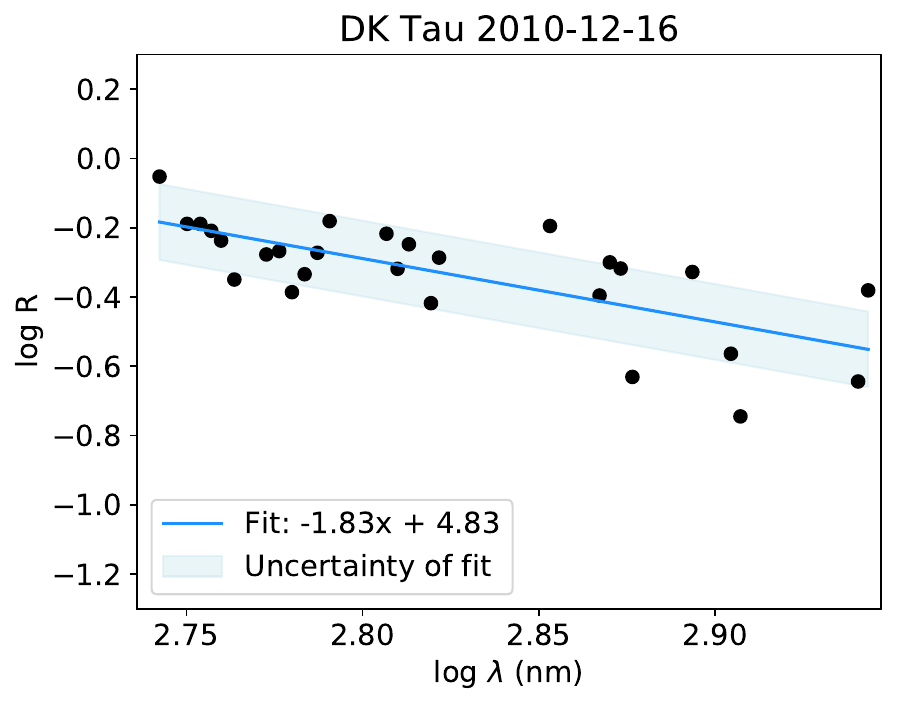}
\includegraphics[scale=0.39]{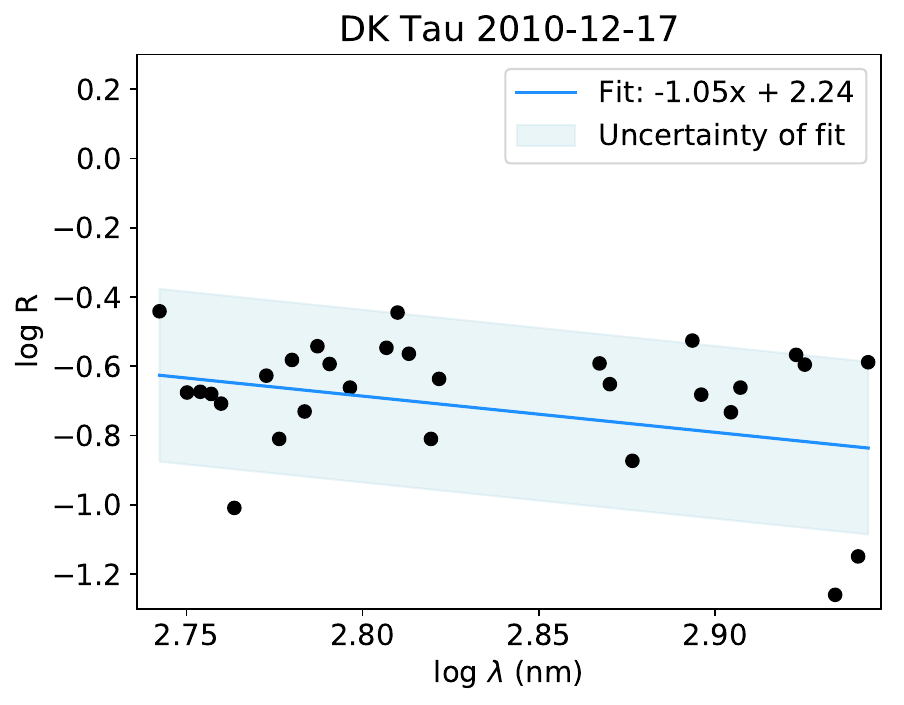}
\includegraphics[scale=0.39]{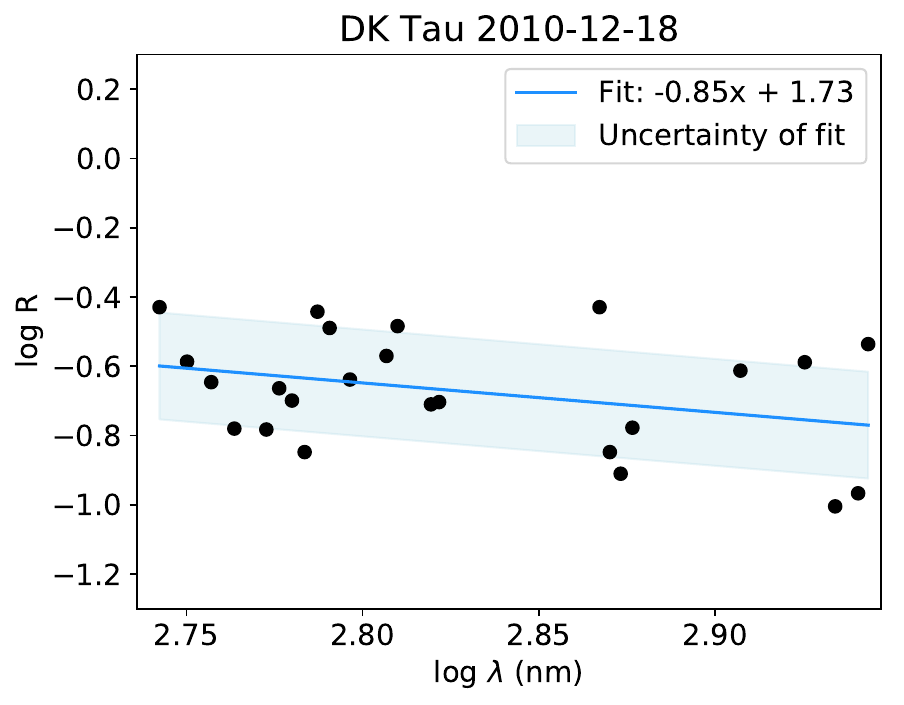}
\includegraphics[scale=0.39]{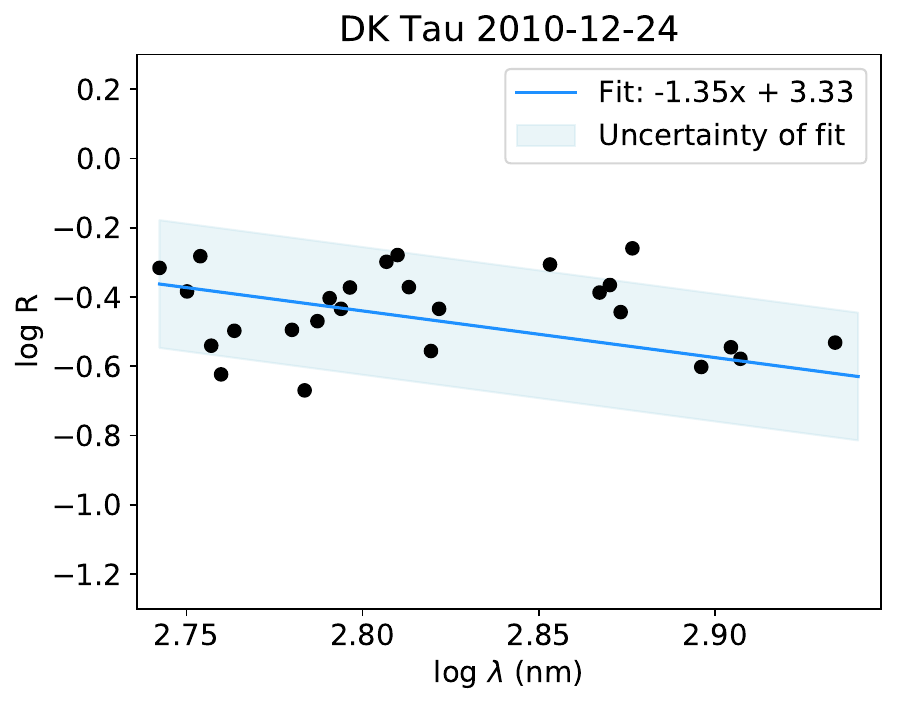}
\includegraphics[scale=0.39]{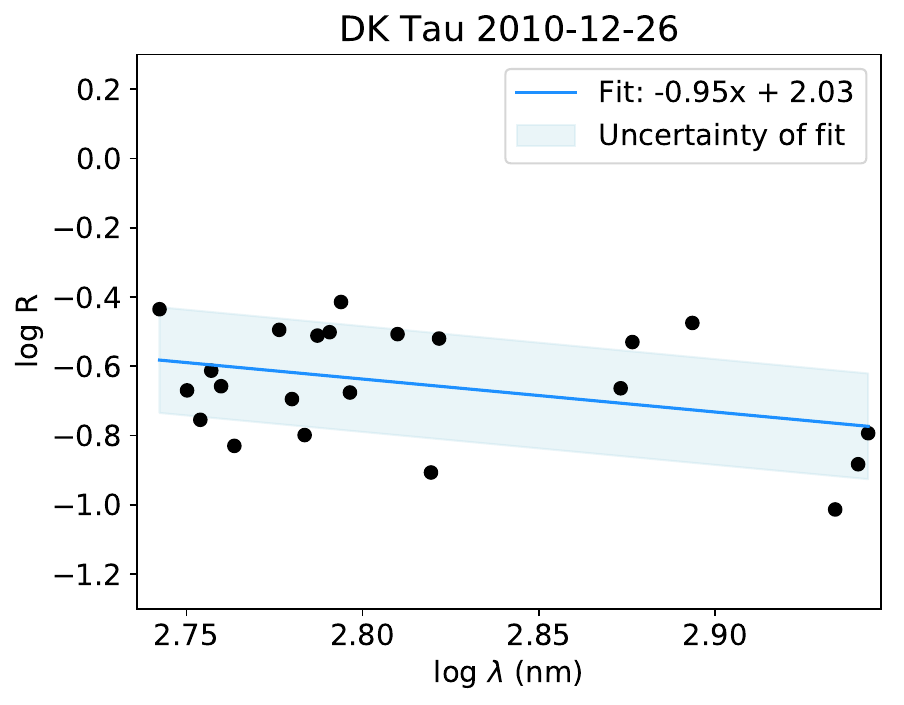}
\includegraphics[scale=0.39]{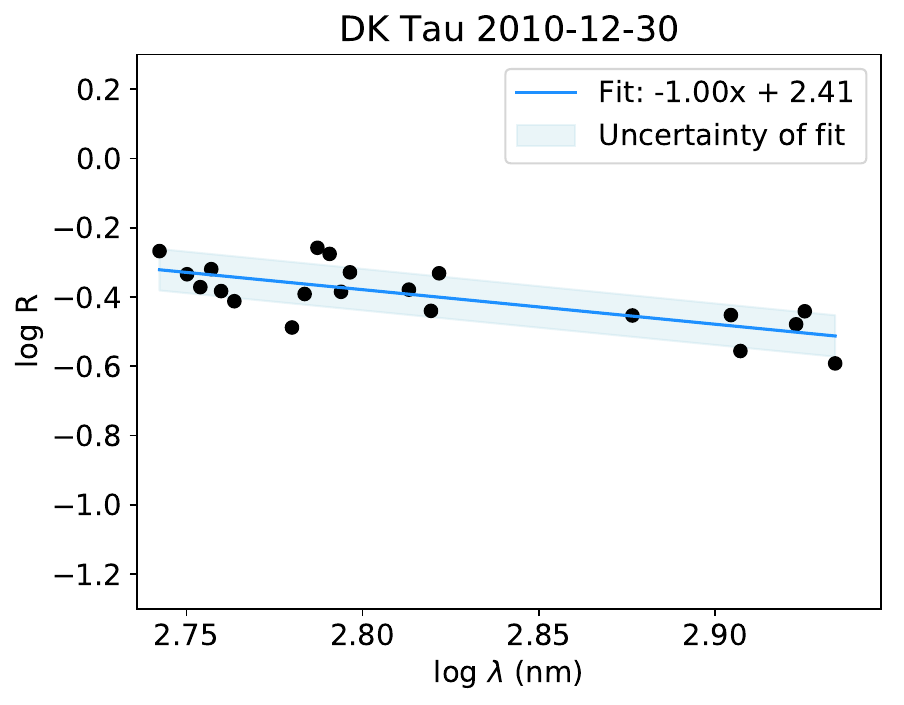}
\caption{Veiling (black dots), the best linear fit (blue line) and the standard deviation (light blue shaded region) as a function of wavelength for the 2010 observations.\label{VeilingLog2010}}
\end{figure*}

\begin{figure*}[hbtp]
\centering
\includegraphics[scale=0.39]{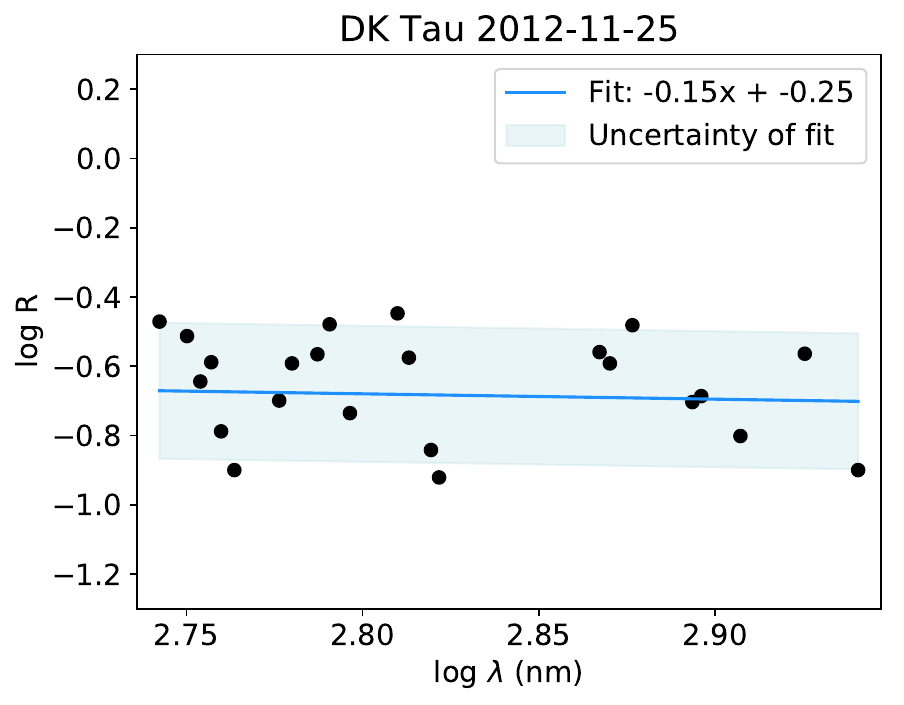}
\includegraphics[scale=0.39]{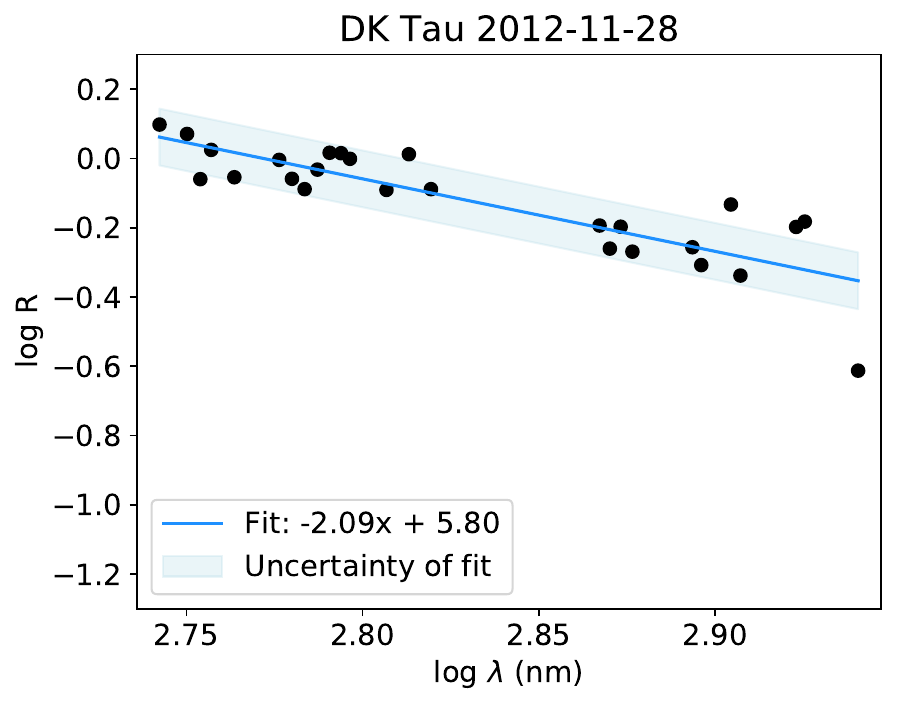}
\includegraphics[scale=0.39]{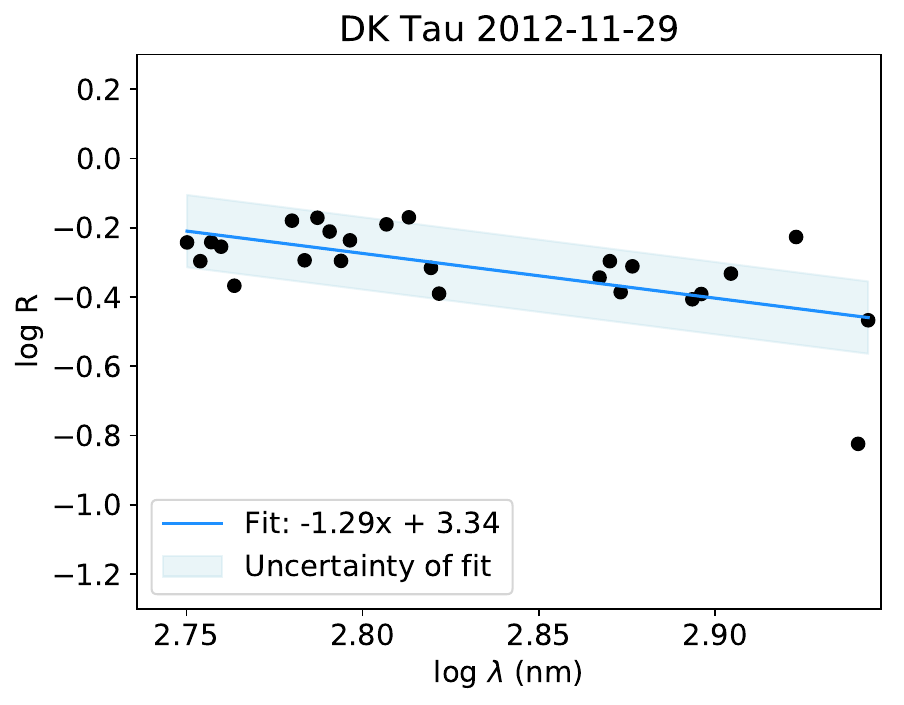}
\includegraphics[scale=0.39]{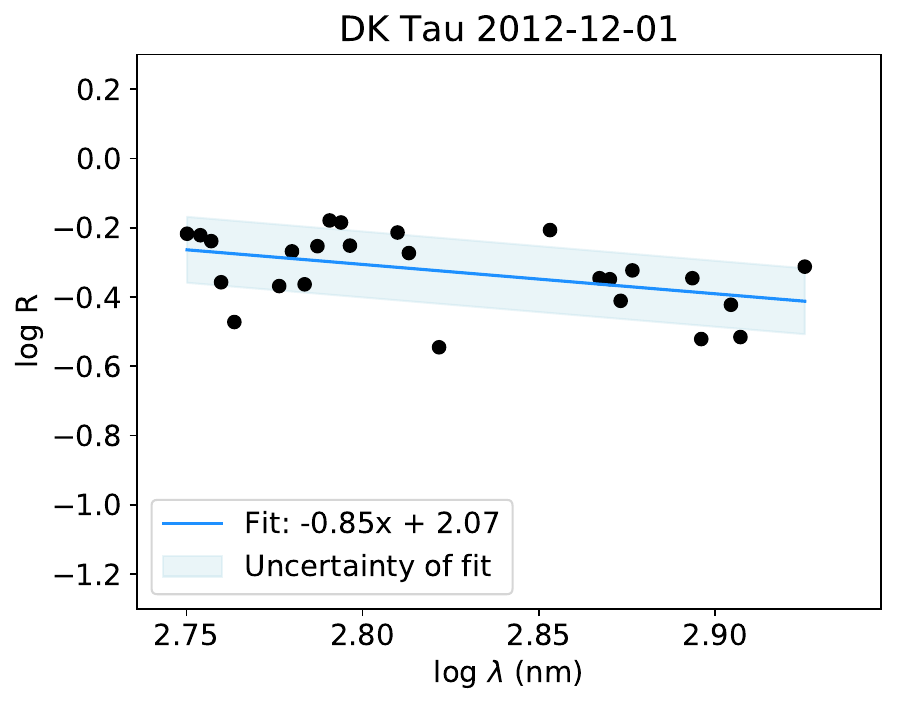}
\includegraphics[scale=0.39]{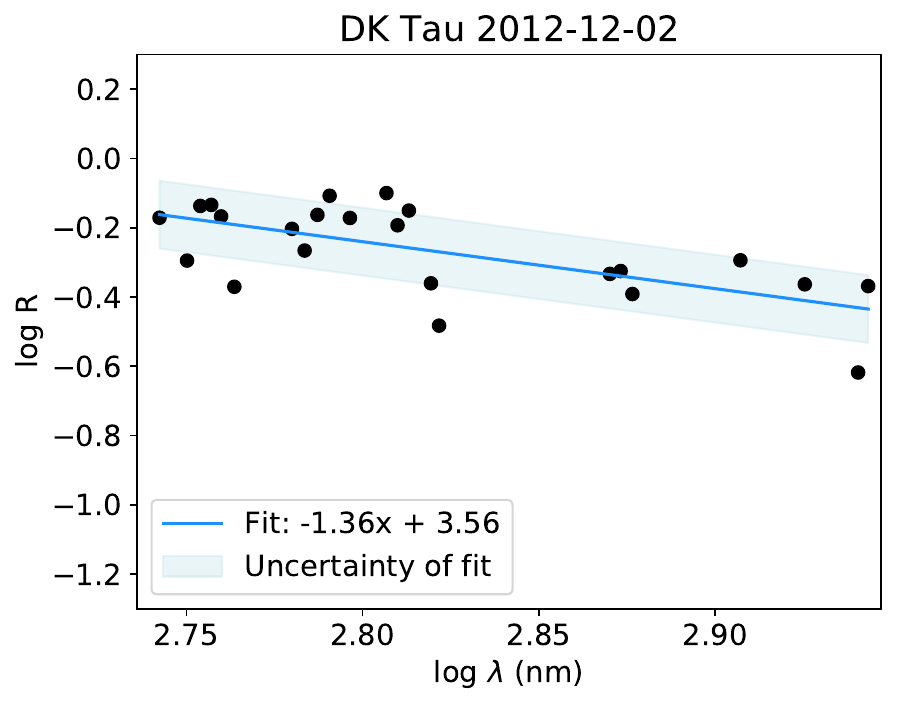}
\includegraphics[scale=0.39]{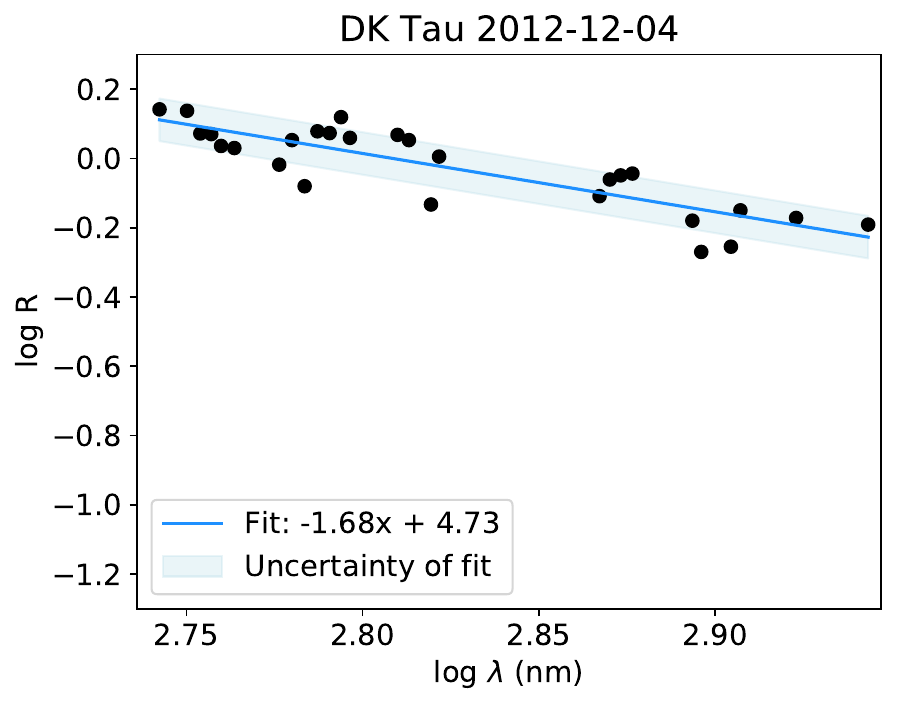}
\includegraphics[scale=0.39]{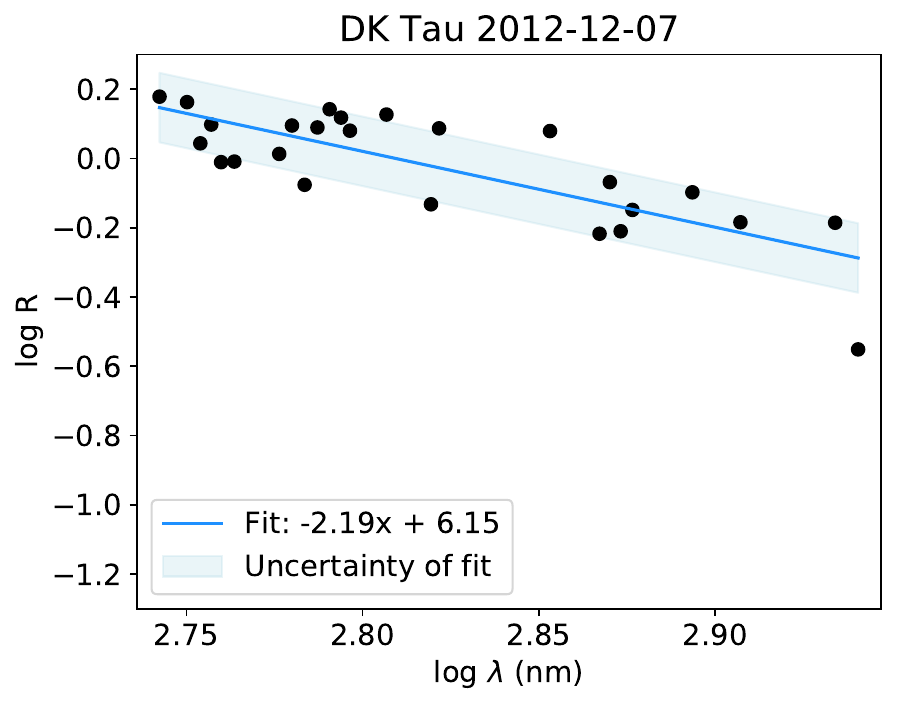}
\includegraphics[scale=0.39]{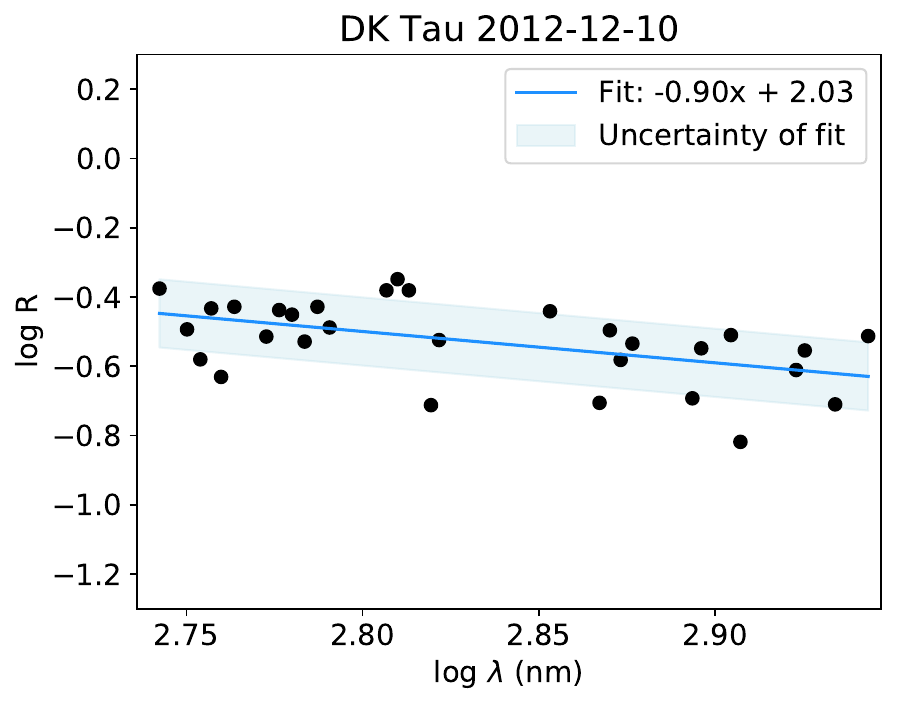}
\includegraphics[scale=0.39]{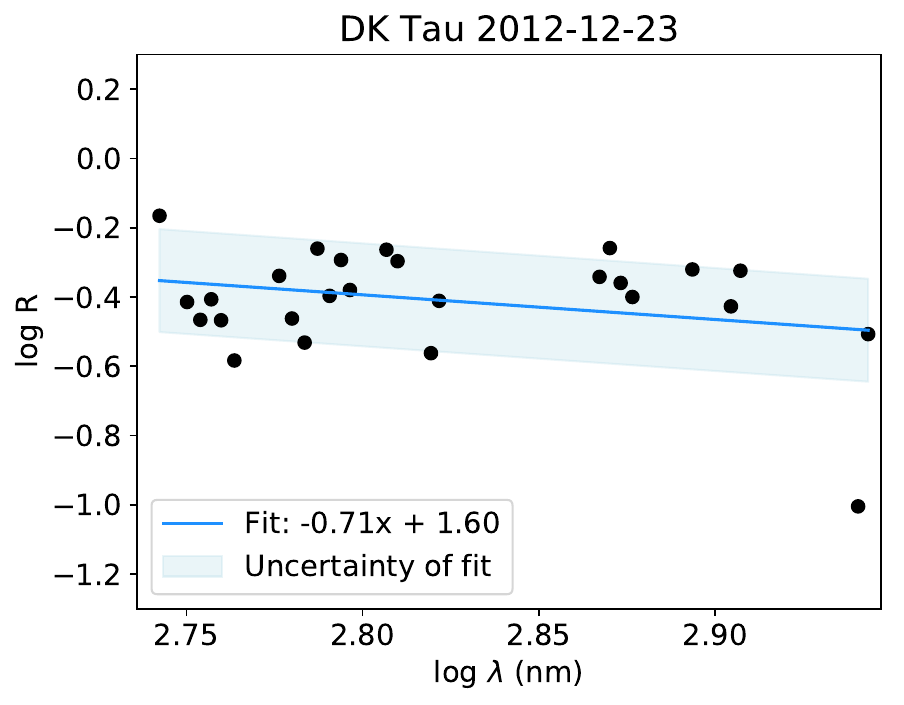}
\caption{Same as Fig.\,\ref{VeilingLog2010}, for the 2012 observations. \label{VeilingLog2012}}
\end{figure*}

\begin{figure}[hbtp]
\centering
\includegraphics[scale=0.55]{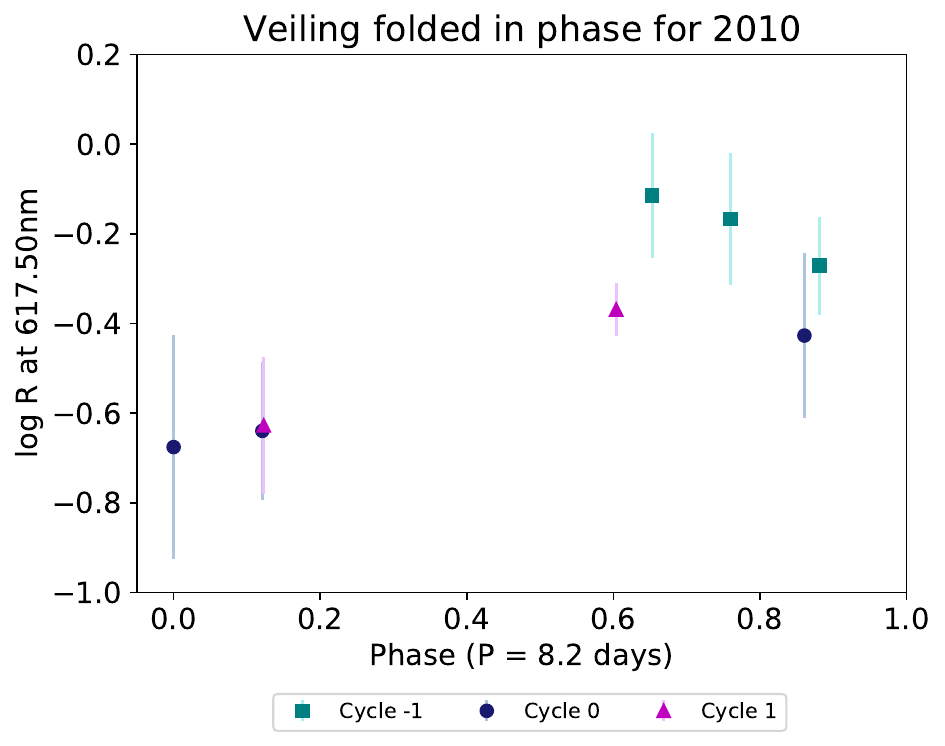}
\includegraphics[scale=0.55]{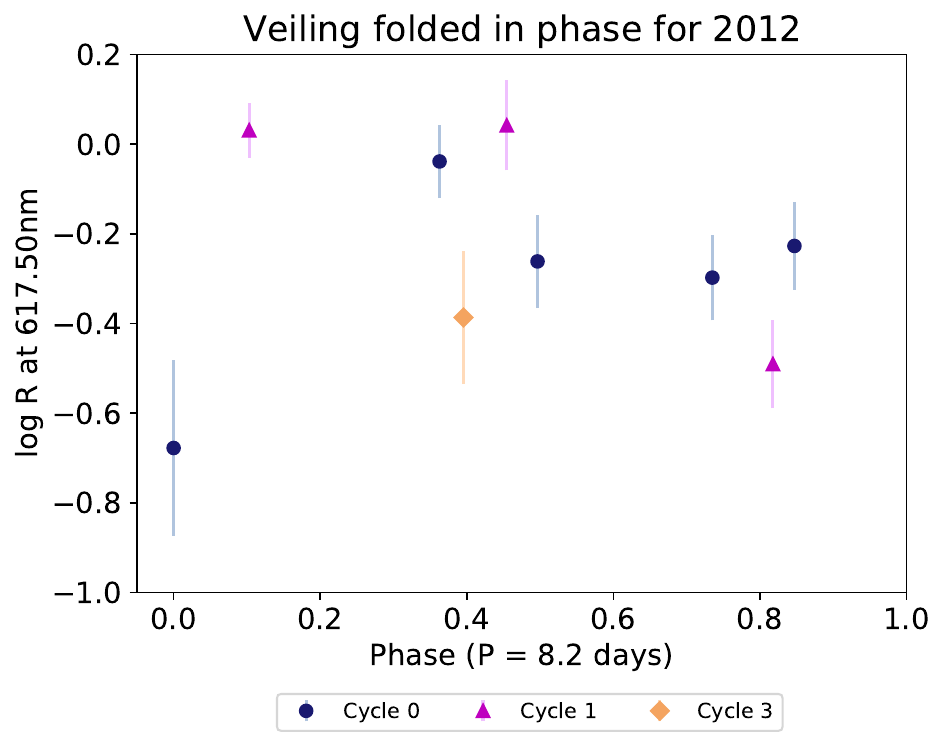}
\caption{Veiling over time, shown folded in phase with an 8.2\,day period, for the 2010 (top panel) and 2012 (bottom panel) data. Cycle 0 for the 2010 epoch starts on December 17, while for the 2012 epoch it starts on the first observation of 2012. This was done in order to display the minimum at the same phase for both epochs. Different colors and symbols represent different rotation cycles. \label{logVeilingTimeFolded}}
\end{figure}

We looked at windows of $\sim$20\,nm throughout the spectra (ranging from 550\,nm to 900\,nm). We then obtained a value of the veiling for the different windows. The window at 617.50\,nm generally showed the best fit between the observation and the template, it is therefore used throughout this work to summarize the veiling for each night (see e.g., Fig.\,\ref{logVeilingTimeFolded}). Once we had values of the veiling as a function of the wavelength, and working in the logarithmic plane, we fit a linear relation through the points (using a least squares polynomial fit). Figures\,\ref{VeilingLog2010} and \ref{VeilingLog2012} show plots of the veiling values (in black) as a function of wavelength for the 2010 and 2012 observations, as well as the best fit ($y = ax + b$ in blue, with $a$ being the slope). The standard deviation is used as the uncertainty of the fit (light blue shaded region). A list of the coefficients of the fits can be found in Appendix\,\ref{VeilingAppendix}. 

We thus studied the variability of veiling for DK\,Tau, not only from night to night, but across our spectra, which cover the optical domain. The peak values of veiling (at $\sim$550\,nm) for each night in 2010 range from 0.2 to 0.9. One observation out of eight has a nearly constant veiling value across its spectrum (i.e., with $a$ < -0.90). For the 2012 observations, the peak values of veiling range from 0.2 to 1.3 (higher than the value from two years before). Three night out of nine have nearly constant veiling values across their spectrum. In general, we can see that when the peak values of veiling are higher, the slope of the fit is steeper (see Fig.\,\ref{VeilingLog2010} and \ref{VeilingLog2012}). Fig.\,\ref{logVeilingTimeFolded} shows the veiling at 617.50\,nm folded in with the stellar rotation phase, for the 2010 and 2012 observations. The error bars are the uncertainty of the fits. We find that the veiling folds well in phase, albeit with more scatter for the 2012 observations. This is likely due to the veiling being modulated by higher accretion activity and not only by the stellar rotation. 

\subsection{Derivation of DK\,Tau's photospheric continuum} \label{PhotCont}

\begin{figure}[hbtp]
\centering
\includegraphics[scale=0.56]{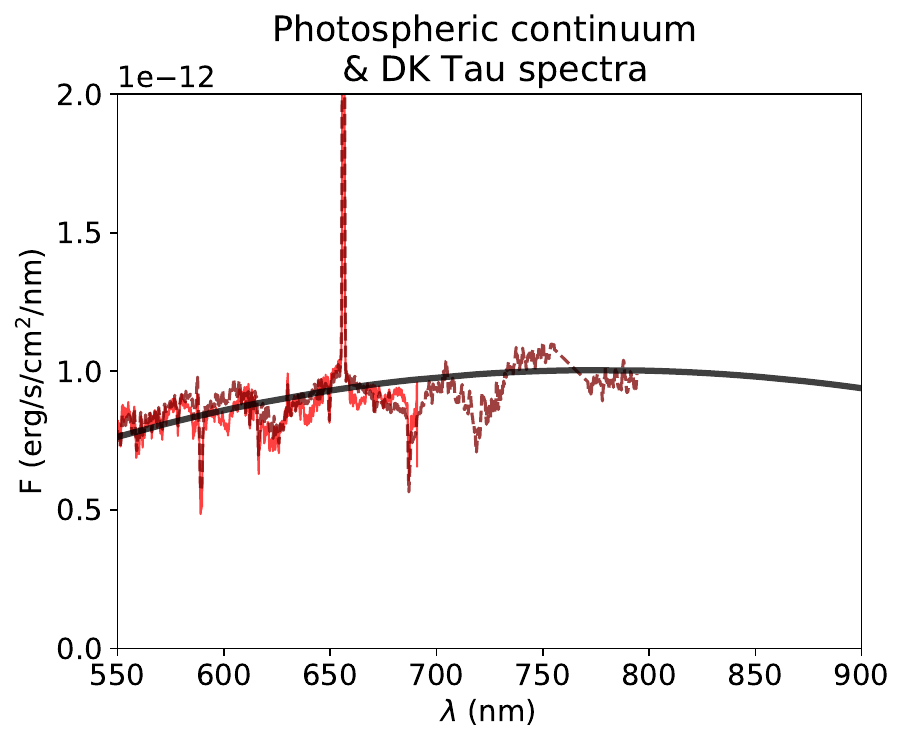}
\caption{DK\,Tau's photospheric continuum as a function of wavelength (thick continuous black line). The thin lines show the flux-calibrated spectra of DK\,Tau taken with HARPS-N (continuous red line) and the  Asiago Telescope (maroon dashed line), both corrected for extinction and deveiled \cite[see][]{2022A&A...667A.124G}. \label{TemplateVsSpec}}
\end{figure}

Because ESPaDOnS spectra are not flux-calibrated, we derived DK\,Tau's photospheric continuum from a high-resolution spectrum  taken at the Telescopio Nazionale Galileo (TNG) using HARPS-N in 2018 by \cite{2022A&A...667A.124G}. The spectrum was corrected for extinction as in  \cite{2022A&A...667A.124G}; we measured the veiling as in the previous section and found a small value with a flat dependence on wavelength in the optical (i.e., R$\sim$ 0.2 with $a$ < -0.90). Fig.\,\ref{TemplateVsSpec} shows the deveiled HARPS-N spectrum. Because it does not cover the entirety of our wavelength range, we also plot a lower resolution spectrum obtained at the Asiago telescope almost simultaneously to the HARPS-N data and used by \cite{2022A&A...667A.124G} for flux-calibration, which extends to slightly longer wavelengths. Finally, to cover the whole region of interest, we used the photospheric continuum of SO879,  a wTTs of K7 spectral type observed with VLT/X-shooter \cite[described in][]{2013A&A...558A.141S}. The X-shooter spectrum was corrected for extinction, then scaled to the distance and luminosity of DK\,Tau, that is $d$ = 132.6\,pc and $L$ = 0.65\,$L_{\odot}$ (with the luminosity value being derived from the flux-calibrated HARPS-N spectrum). As the SO879 continuum proved to be a good description of the DK\,Tau continuum in the spectral range where they overlap, we adopt it as a good representation of the photospheric continuum over the whole range of the ESPaDOnS spectra discussed in the following. 

In summary then to determine the values of veiling (see Sect.\,\ref{Veiling}), we used an ESPaDOnS spectrum of the wTTs TAP45 because of its high resolution. As ESPaDOnS spectra are not flux-calibrated, we could not use that spectrum to extract DK\,Tau's photospheric continuum. Conversely, the X-shooter spectrum of SO879 could not be used to derive the veiling because of its lower resolution. In the absence of an appropriate wTTs spectrum to use as a template in both circumstances, we worked with TAP45 and SO879. 

\section{Results} \label{sec:results}

\subsection{Accretion luminosity from emission lines} \label{Lacc}

\begin{figure}[hbtp]
\centering
\includegraphics[scale=0.55]{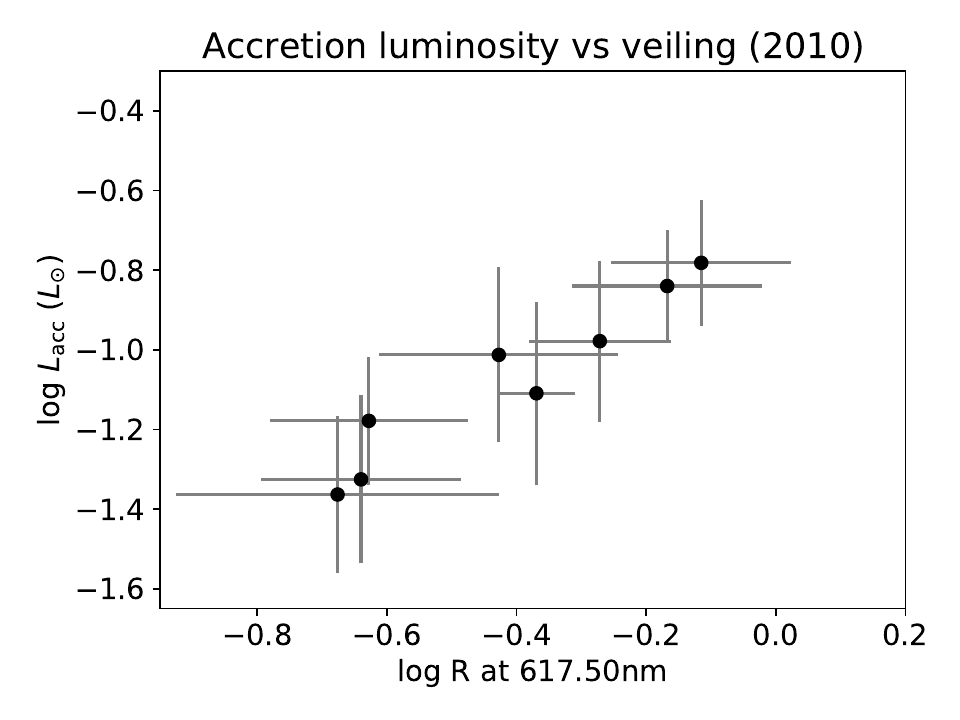}
\includegraphics[scale=0.55]{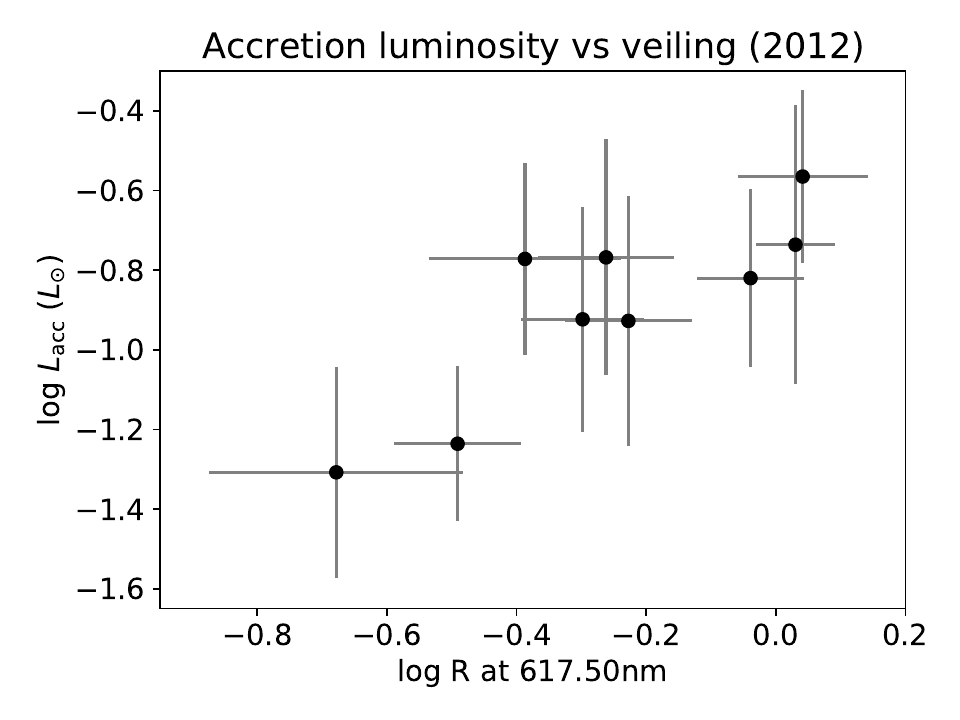}
\caption{Accretion luminosity as a function of the veiling at 617.50\,nm, for the 2010 (top panel) and 2012 (bottom panel) data. \label{logLacc_vs_logVeiling}}
\end{figure}

In order to calculate the accretion luminosities for each night, we first measured the equivalent width (EW) of several accretion-powered emission lines, namely H$\alpha$, H$\beta$, H$\gamma$, the He{\sc i} lines at 447.1\,nm, 501.6\,nm, 587.6\,nm, 667.8\,nm and 706.5\,nm, as well as the Ca{\sc ii} infrared triplet at 849.8\,nm, 854.2\,nm and 866.2\,nm. We corrected the EWs for veiling, then we flux-calibrated them using our template of DK\,Tau's photospheric continuum to obtain line fluxes, before converting the line fluxes into line luminosities. Finally, using the empirical relations in Table\,B.1. from \cite{2017A&A...600A..20A}, we obtained the values of the accretion luminosities for each night and for each line. Plots of the accretion luminosities on a nightly basis can be found in Appendix\,\ref{LaccAppendix}. Figure\,\ref{logLacc_Phase} shows accretion luminosity values averaged over the different lines and folded in phase. We calculated the weighted average, and used the weighted standard deviation of the spread in the values found from the different lines as the error bars. The values for each night are listed in Table\,\ref{TableMdot}. We can see that the accretion luminosity changes with time by a factor of up to approximately 6. It changes in phase with the stellar rotation in the 2010 epoch data, but less clearly in the 2012 data. In Fig.\,\ref{logLacc_vs_logVeiling}, we can see that the accretion luminosity correlates with the veiling in an almost linear fashion (see Sect.\,\ref{LaccComp} for a discussion of the correlation). 

\begin{figure}[hbtp]
\centering
\includegraphics[scale=0.56]{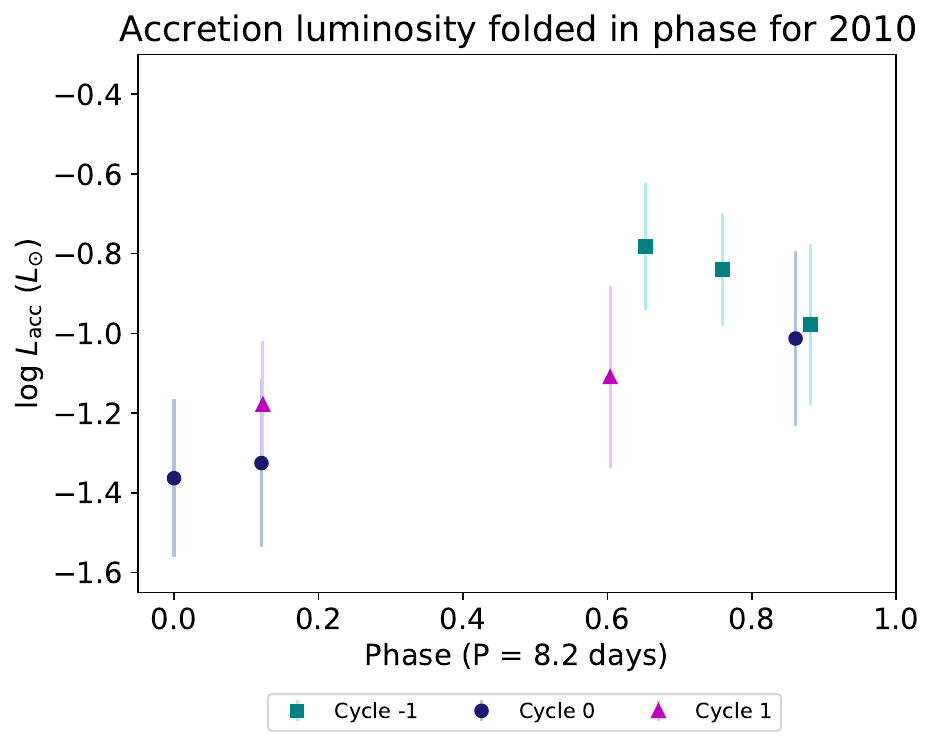}
\includegraphics[scale=0.56]{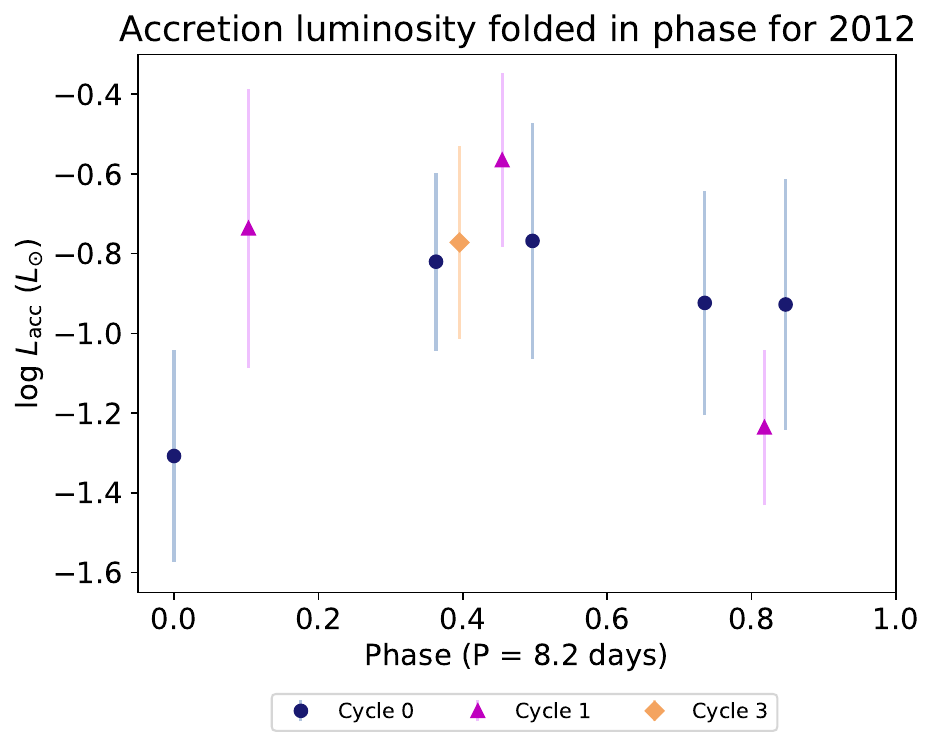}
\caption{Accretion luminosity over time, shown folded in phase with an 8.2\,day period, for the 2010 (top panel) and 2012 (bottom panel) data. Cycle 0 for the 2010 epoch starts on December 17 2010, while for the 2012 epoch it starts on the first observation of 2012, as in previous figures. Different colors and symbols represent different rotation cycles. \label{logLacc_Phase}}
\end{figure}

\begin{figure}[hbtp]
\centering
\includegraphics[scale=0.55]{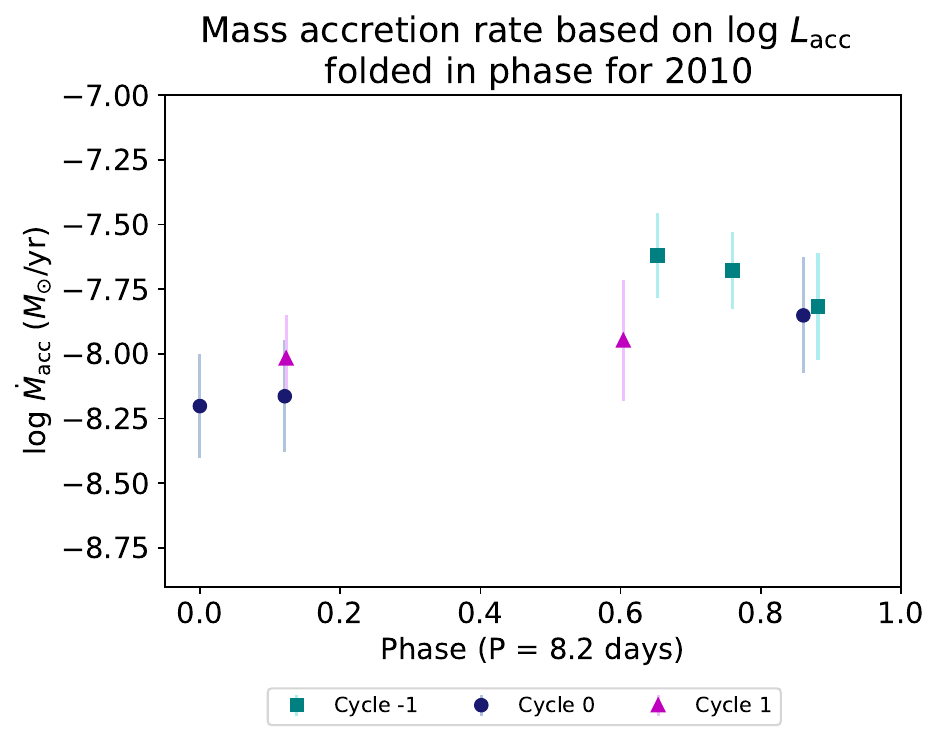}
\includegraphics[scale=0.55]{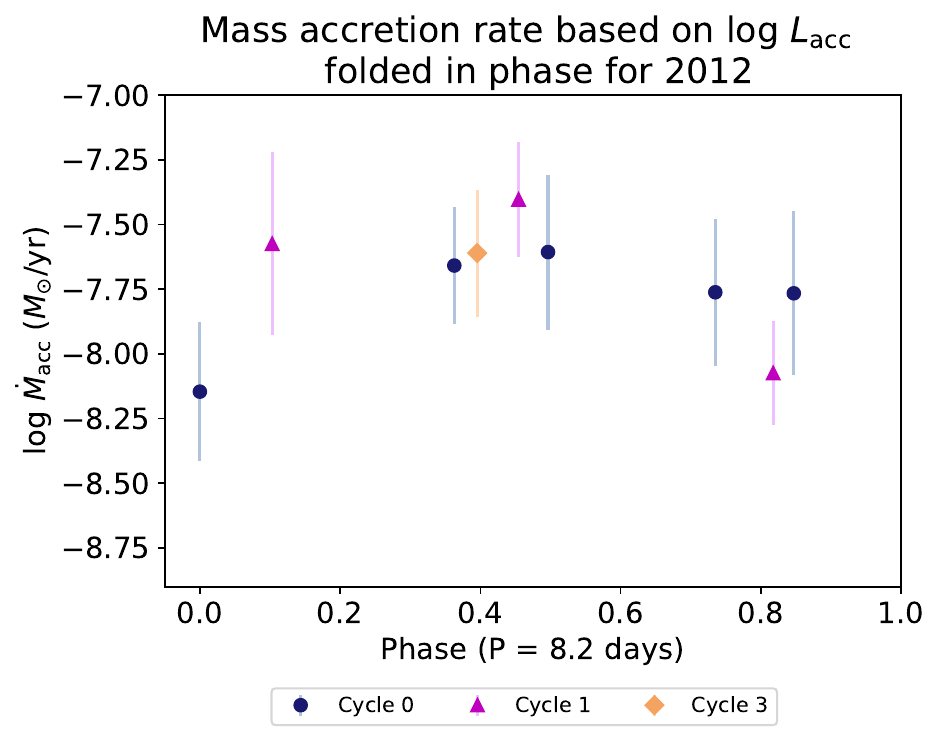}
\caption{Mass accretion rate based on the accretion luminosity over time, shown folded in phase with an 8.2\,day period, for the 2010 (top panel) and 2012 (bottom panel) data. Cycle 0 for the 2010 epoch starts on December 17, while for the 2012 epoch it starts on the first observation of 2012, as in previous figures. Different colors and symbols represent different rotation cycles.  \label{logMacc_via_Lacc_phase}}
\end{figure}

Next, we calculated the mass accretion rate onto the star for each night using the values of the accretion luminosities. The conversion was made using equation 8 from \cite{1998ApJ...492..323G}: 
\begin{equation}
L_{\text{acc}} \simeq \frac{G M_\star \dot{M}_{\text{acc}}}{R_\star} \left( 1 - \frac{R_\star}{R_{\text{in}}} \right)
\end{equation}
with $R_\star$ = 2.48\,$R_{\odot}$, $M_\star$ = 0.7\,$M_{\odot}$ and  $R_{\text{in}}$ = 5\,$R_{\star}$ (as is typically used). Figure\,\ref{logMacc_via_Lacc_phase} shows the mass accretion rate obtained via $L_{\text{acc}}$ and folded in phase. As the mass accretion rate is proportional to the accretion luminosity, we take the error bars on $\dot{M}_{\text{acc}}$ as being the same as the ones on $L_{\text{acc}}$, which are the dominating components, ignoring the error bars on $M_\star$ and $R_\star$. Furthermore, we are focusing on the variability and the latter do not change from night to night. 

For the accretion luminosity, we find values ranging from log\,($L_{\text{acc}}$[$L_\odot$]) = -1.36 to log\,($L_{\text{acc}}$[$L_\odot$]) = -0.57. This leads to mass accretion rate values ranging from log\,($\dot{M}_{\text{acc}}$[$M_\odot$\,yr$^{-1}$]) = -8.20 to log\,($\dot{M}_{\text{acc}}$[$M_\odot$\,yr$^{-1}$]) = -7.40. The values of the accretion luminosities and mass accretion rates that we find for each night are listed in Table\,\ref{TableMdot}. In comparison, \cite{2011ApJ...730...73F} find log\,($\dot{M}_{\text{acc}}$[$M_\odot$\,yr$^{-1}$]) = -7.40, a value that is in agreement with the higher end of our range. \cite{1998ApJ...492..323G} and \cite{2013ApJ...767..112I} quote similar values of log\,($\dot{M}_{\text{acc}}$[$M_\odot$\,yr$^{-1}$]) = -7.42 and -7.47. Our results are also in agreement with the value quoted by \cite{2018ApJ...868...28F} of log\,($\dot{M}_{\text{acc}}$[$M_\odot$\,yr$^{-1}$]) = -7.86, a value approximately in the middle of our range. \cite{2022A&A...667A.124G} find log\,($\dot{M}_{\text{acc}}$[$M_\odot$\,yr$^{-1}$]) = -8.33. This agrees within the error bars of the lower end of our range. 


\subsection{Accretion shock models} \label{ShMod}

The emission of accretion shocks has been modeled starting with \cite{1998ApJ...509..802C}. What is observed from the accretion shock is determined by the energy of the shock and by the size of the accretion spot on the stellar surface. These accretion shock models therefore depend on the accretion energy flux $\mathscr{F}$, the fractional surface coverage of the accretion spots $f$ and on the stellar properties. The models adopt a one-dimensional plane-parallel geometry for the accretion column and do not distinguish between a single accretion spot or a multitude of spots.

The accretion energy flux is the flux of energy that the accretion column carries into the shock, and it is defined by equation 10 from \cite{1998ApJ...509..802C}: 
\begin{equation}
\begin{aligned}
\mathscr{F} = \frac{1}{2} \, \rho \, v_{s}^{3}
\end{aligned}
\end{equation}
where $\rho$ is the density of the material in the accretion column and $v_{s}$ the free-fall velocity of that material. It is assumed that the velocity is constant, as it depends on the mass and radius of the star via equation 1 from \cite{1998ApJ...509..802C}:
\begin{equation}
\begin{aligned}
v_{s}^{3} = \left| \frac{2G M_\star}{R_\star} \right|^{1/2} \, \left| 1- \frac{R_\star}{R_{\text{i}}} \right|^{1/2}
\end{aligned}
\end{equation}
where $M_\star$ is the stellar mass, $R_\star$ is the stellar radius and $R_{\text{i}}$ is the radius where the magnetosphere truncates the circumstellar disk and is assumed to be 5\,$R_\star$. It follows that low or high values of $\mathscr{F}$ represent low or high-density columns. 

Each energy flux is scaled by a filling factor $f$, representing the fraction of the stellar surface covered by accretion spots. Its value varies between 0 and 1: when $f$ = 0, the spot is non-existent; when $f$ = 1, the spot has the size of the star; when, for example, $f$ = 0.1, this indicates that 10\,$\%$ of the stellar surface is covered by accretion spots (regardless of it being a single large accretion spot or several smaller spots). A decrease in the filling factor will downscale the emission of the accretion column independently of wavelength, while a change in the energy flux will shift the peak of the emission in wavelength. 

The shock models that we worked with were made using DK\,Tau's parameters: a distance $d$ = 132.6\,pc, a stellar luminosity $L$ = 0.65\,$L_{\odot}$, a radius $R_\star$ = 2.48\,$R_{\odot}$, an effective temperature $T_{\textrm{eff}}$ = 4\,000\,K and a mass $M_\star$ = 0.7\,$M_{\odot}$. They were computed for different energy fluxes, namely $\mathscr{F}$\,=\,$1\,\times\,10^{9}$\,erg\,s$^{-1}$\,cm$^{-2}$, $\mathscr{F}$\,=\,$3\,\times\,10^{9}$\,erg\,s$^{-1}$\,cm$^{-2}$, $\mathscr{F}$\,=\,$1\,\times\,10^{10}$\,erg\,s$^{-1}$\,cm$^{-2}$, $\mathscr{F}$\,=\,$3\,\times\,10^{10}$\,erg\,s$^{-1}$\,cm$^{-2}$, $\mathscr{F}$\,=\,$1\,\times\,10^{11}$\,erg\,s$^{-1}$\,cm$^{-2}$ and $\mathscr{F}$\,=\,$3\,\times\,10^{11}$\,erg\,s$^{-1}$\,cm$^{-2}$. 

\begin{figure}[hbtp]
\centering
\includegraphics[scale=0.56]{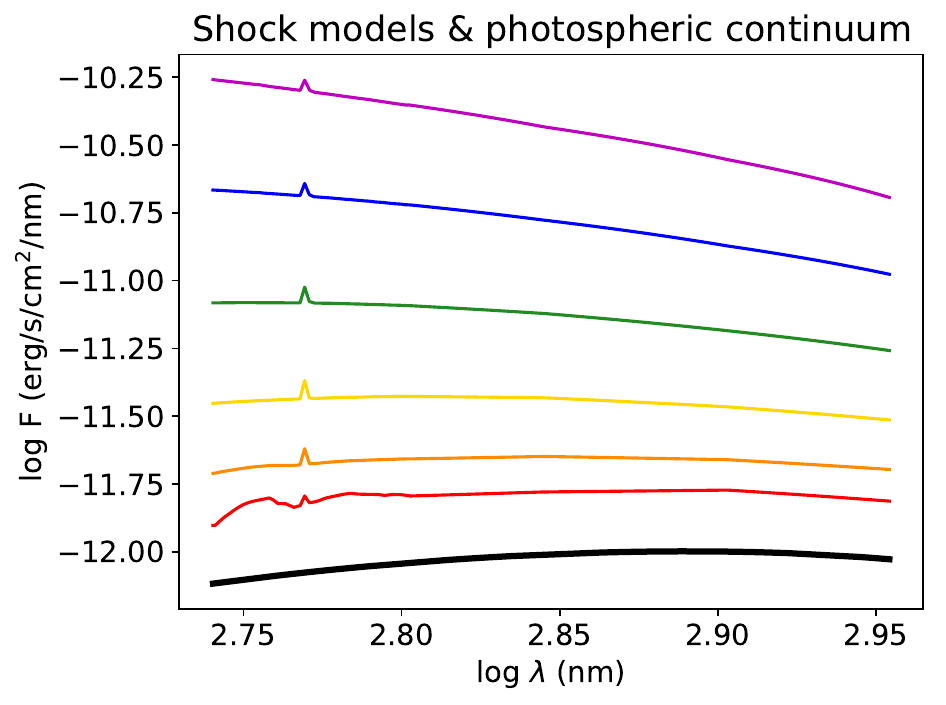}
\includegraphics[scale=0.56]{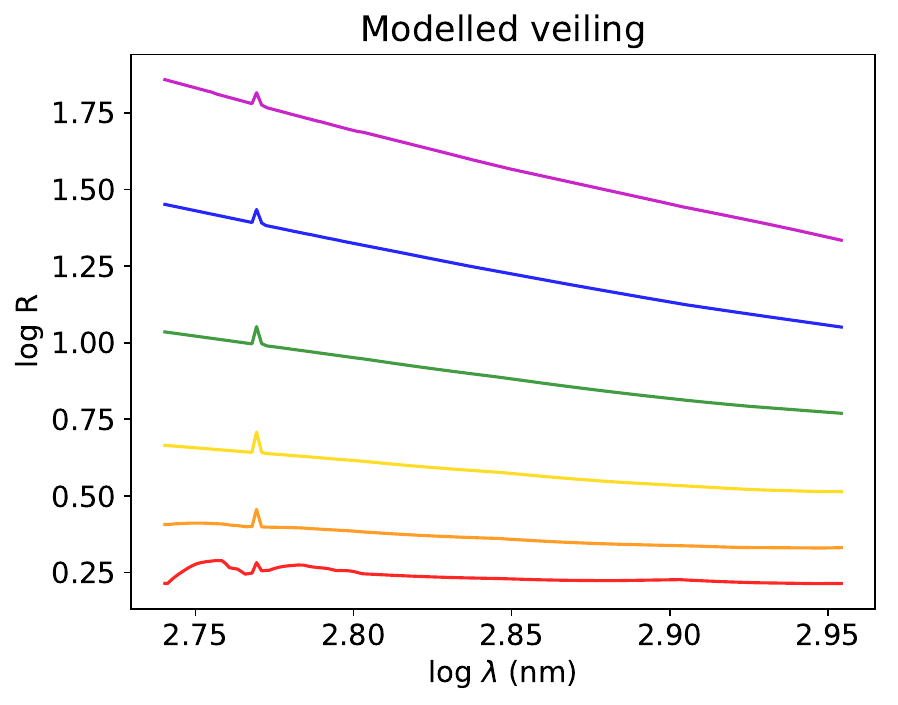}
\includegraphics[scale=0.23]{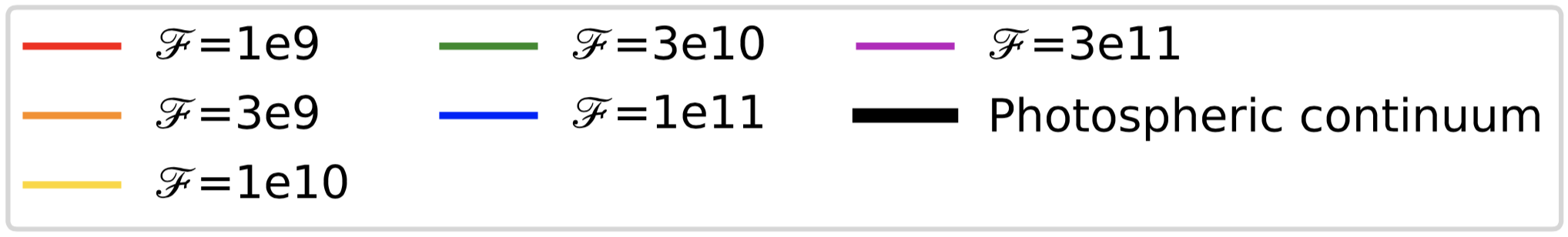}
\caption{Models of the excess flux from accretion shocks (in rainbow colors) and photospheric continuum (in black) as a function of wavelength (top panel). Modeled veiling (in rainbow colors) computed from the different shock models and the photospheric continuum (bottom panel). \label{ModelsTemplate_Veiling}}
\end{figure}

The top panel of Fig.\,\ref{ModelsTemplate_Veiling} shows the shock models for different energy fluxes, as well as DK\,Tau's photospheric continuum. We calculated the ratio between the different modeled accretion shock fluxes and the photospheric continuum flux, in order to obtain different values of model-predicted veiling as a function of wavelength (see bottom panel of Fig.\,\ref{ModelsTemplate_Veiling}). We can see that, in the optical range, the slope of the model-predicted veiling increases monotonically with the energy flux $\mathscr{F}$. Therefore, the slope of the modeled veiling seems to univocally characterize the accretion energy flux. This allows us to assign a specific value of $\mathscr{F}$ to a specific slope in the observed veiling. 

\begin{figure}[hbtp]
\centering
\includegraphics[scale=0.58]{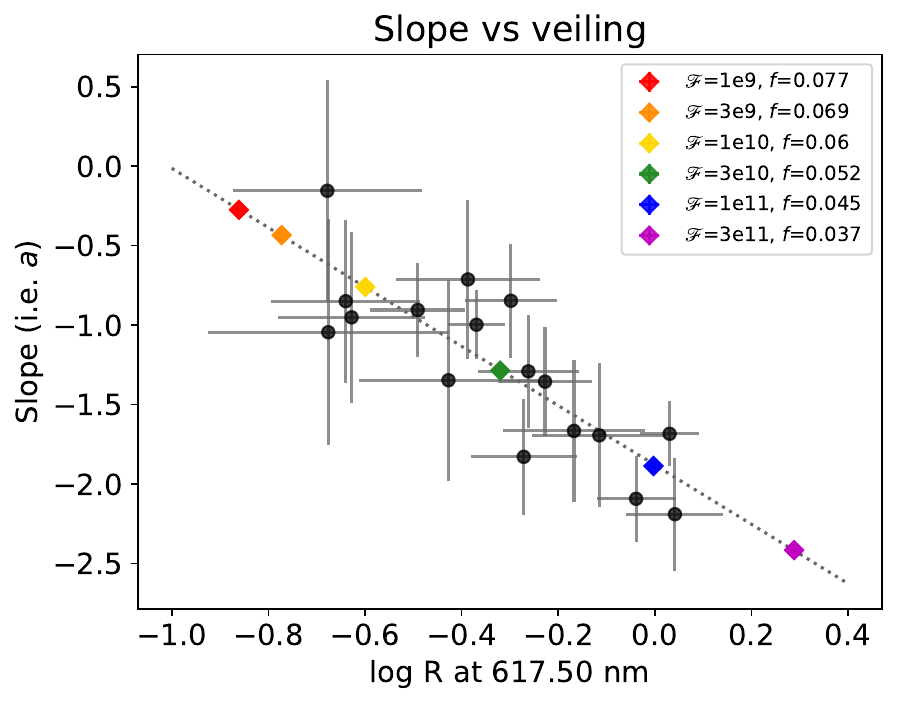}
\caption{Slope of the veiling for the different models (diamonds in rainbow colors), and for the 2010 and 2012 observations (black dots) as a function of the veiling value at 617.50\,nm. The vertical error bars are the errors on $a$, the horizontal ones are the uncertainty of the fits. The gray dotted line highlights the trend followed by the observations.\label{ModelsObservationsVeiling}}
\end{figure}

Working in the logarithmic plane, we computed the slopes of the observed and modeled veiling values by fitting the points with a linear relation: $y = ax + b$. We then plotted the slopes (i.e., $a$) as a function of the value of the veiling at 617.50\,nm in Fig.\,\ref{ModelsObservationsVeiling}. We linearly interpolated the values of the slopes, of the veiling at 617.50\,nm, and of the energy fluxes between the six models that we used (but only show these six models in Fig.\,\ref{ModelsObservationsVeiling} for clarity). 

The observed veiling values follow the aforementioned trend of a higher slope with higher veiling values (highlighted by the gray dotted line). Regarding the model-predicted veiling values, they are scaled by the filling factors which are free parameters. Because, in the optical range, the slopes of the modeled veiling are only dependent on the value of the energy flux and independent of the filling factors, we first derive the best value of the energy flux for each observation, by matching the slope of the modeled veiling with the slope of the observational veiling (i.e., matching along the y-axis). Next, adjusting the values of the filling factors shifts the modeled veiling values along the x-axis only. This allows us to align the models with the observations by varying the filling factors. We thus derive the energy flux and the appropriate filling factor for that energy flux for each night. Two examples of the overlap of the observed and modeled veiling can be found in Appendix\,\ref{ShModAppendix}. 

In the literature \cite[see e.g.,][]{2013ApJ...767..112I, 2019ApJ...874..129R, 2022AJ....163..114E, 2022AJ....164..201P}, it is usually a combination of several energy fluxes, each with its own filling factor, that are used to fit a single observation, with each energy flux better constrained by different regions of the spectrum. In general, the higher energy fluxes (corresponding to higher densities of the material in the accretion column) peak in the UV range and the lower energy fluxes peak in the optical. In this work, however, as we have a narrower wavelength range (i.e., 550\,nm - 900\,nm) that excludes the UV, we are using a single energy flux with its assigned filling factor. Nevertheless, we are able to constrain the models to our data adequately. We are thus making the approximation that, for a given night, one homogeneous accretion spot, characterized by a single value of $\mathscr{F}$ and of $f$, is dominating the optical emission. 

We find energy fluxes ranging from 1.00\,$\times$\,$10^{9}$ to 2.15\,$\times$\,$10^{11}$\,erg\,s$^{-1}$\,cm$^{-2}$. For the different energy fluxes, we find filling factors ranging from 0.026 to 0.117, implying that the accretion columns cover from 2.6\% to 11.7\% of the stellar surface. These numbers are comparable to the ones listed in the literature for other cTTs, though it should be noted that they use a multicolumn approach, whereas we assume a single column and spot. For example, using three accretion columns with $\mathscr{F}$\,=\,$1\,\times\,10^{10}$\,erg\,s$^{-1}$\,cm$^{-2}$, $\mathscr{F}$\,=\,$1\,\times\,10^{11}$\,erg\,s$^{-1}$\,cm$^{-2}$, $\mathscr{F}$\,=\,$1\,\times\,10^{12}$\,erg\,s$^{-1}$\,cm$^{-2}$, \cite{2019ApJ...874..129R} mention filling factors ranging from 5.00\,$\times$\,$10^{-5}$ to 0.39 for various low-mass cTTs (namely DM\,Tau, GM\,Aur, SZ\,45, TW\,Hya and VW\,Cha). For the same energy fluxes, \cite{2021Natur.597...41E} find filling factors ranging from 5.30\,$\times$\,$10^{-5}$ to 0.18 for the cTTs GM\,Aur. For the same energy fluxes once more, \cite{2022AJ....164..201P} mention filling factors ranging from 4.85\,$\times$\,$10^{-5}$ to 0.23 for several cTTs (namely CVSO\,58, CVSO\,90, CVSO\,104, CVSO\,107, CVSO\,109A, CVSO\,146, CVSO\,165A, CVSO\,165B and CVSO\,176). 

The values of the energy fluxes and filling factors that we find for each night are listed in Table\,\ref{TableMdot}. We can see that they are anticorrelated: higher energy fluxes are paired with smaller filling factors. Conversely, filling factors increase when the energy fluxes decrease, in order to match the observed veiling. This is to be expected for a constant accretion rate, as the product of the energy flux by the filling factor is proportional to the mass accretion rate \cite[see equation 11 from][]{1998ApJ...509..802C}. In other words, in order to generate the same amount of flux, lower energy fluxes need to be paired with larger filling factors \cite[see e.g.,][]{1998ApJ...509..802C, 2013ApJ...767..112I}. 

\begin{figure}[hbtp]
\centering
\includegraphics[scale=0.56]{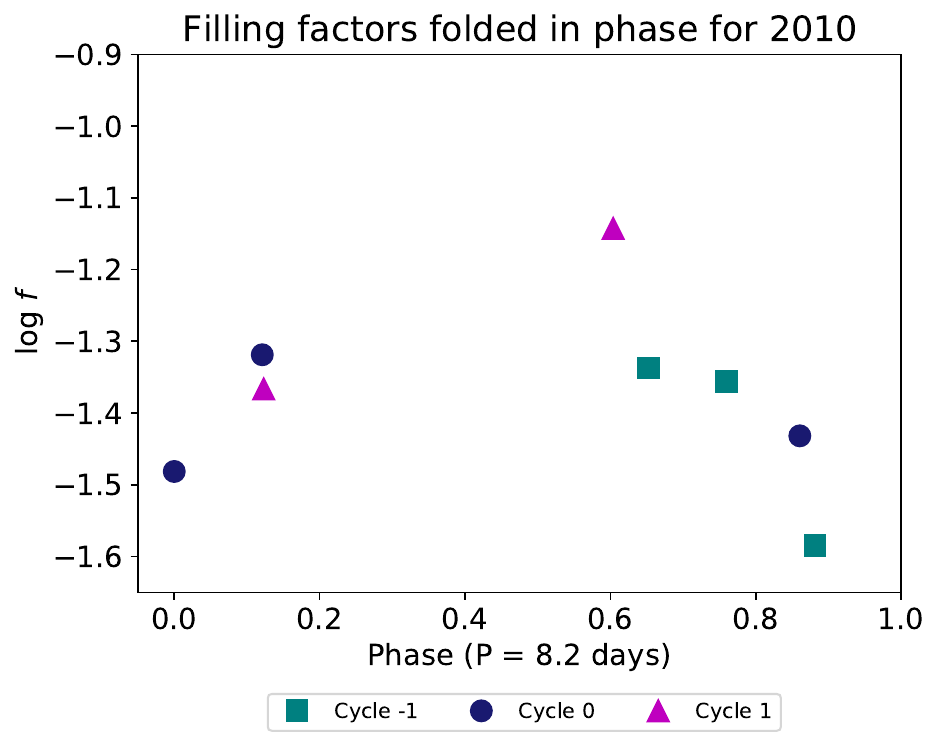}
\includegraphics[scale=0.56]{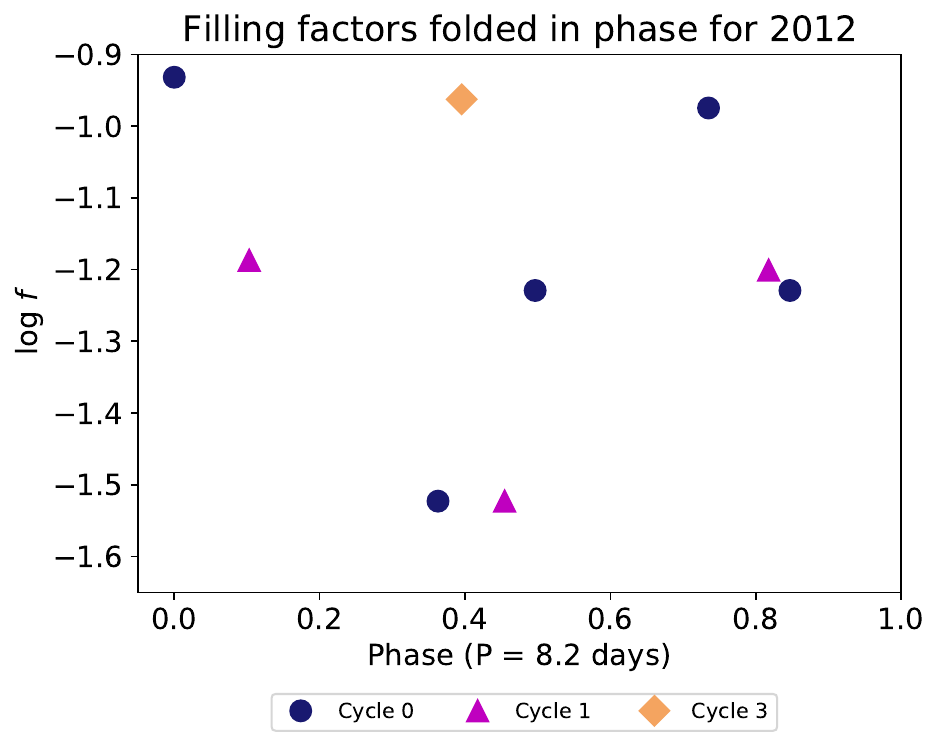}
\caption{Filling factors over time, shown folded in phase with an 8.2\,day period, for the 2010 (top panel) and 2012 (bottom panel) data. Cycle 0 for the 2010 epoch starts on December 17, while for the 2012 epoch it starts on the first observation of 2012, as in previous figures. Different colors and symbols represent different rotation cycles. \label{ff_Phase}}
\end{figure}

\begin{figure}[hbtp]
\centering
\includegraphics[scale=0.56]{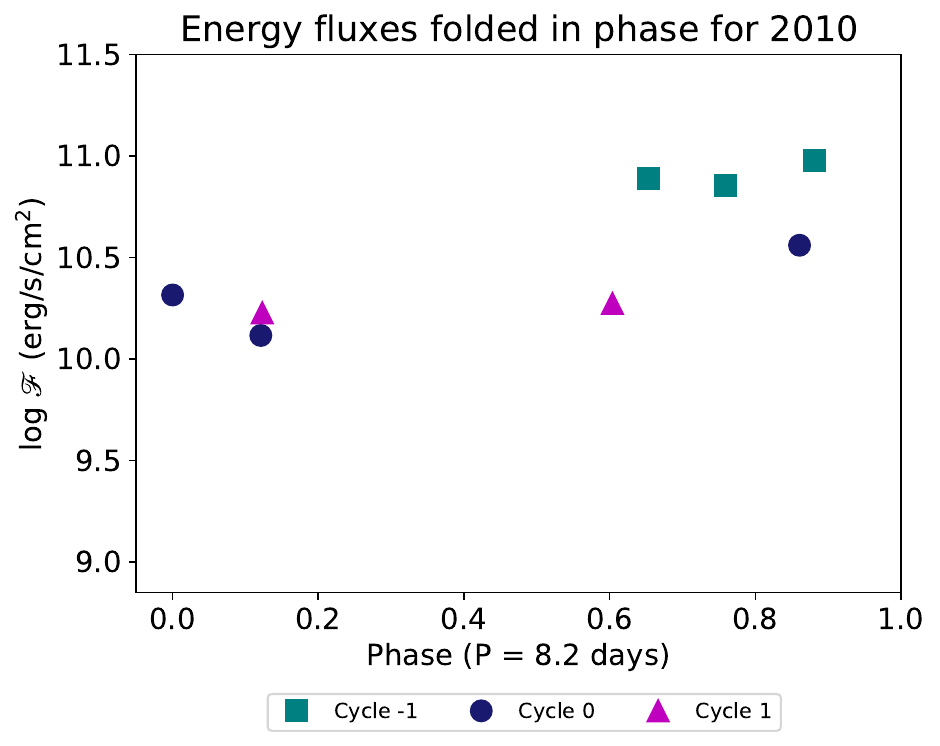}
\includegraphics[scale=0.56]{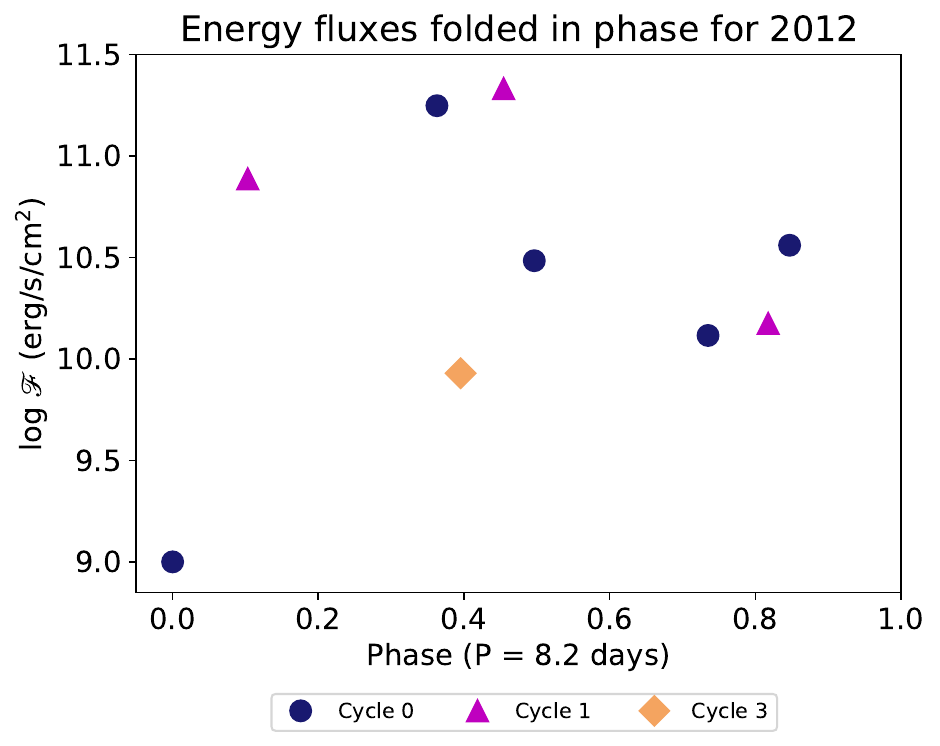}
\caption{Energy fluxes over time, shown folded in phase with an 8.2\,day period, for the 2010 (top panel) and 2012 (bottom panel) data. Cycle 0 for the 2010 epoch starts on December 17, while for the 2012 epoch it starts on the first observation of 2012, as in previous figures. Different colors and symbols represent different rotation cycles. \label{F_Phase}}
\end{figure}

Figure\,\ref{ff_Phase} shows the filling factors folded in phase, while Fig.\,\ref{F_Phase} shows the energy fluxes folded in phase. We can see that both quantities vary in phase with the stellar rotation, notably in the 2010 epoch data. 

Next, using the values of energy fluxes and filling factors that we obtained, and using equation 11 from \cite{1998ApJ...509..802C}:
\begin{equation}
\begin{aligned}
\mathscr{F} = 9.8 \times 10^{10} \, & \text{ergs} \, \text{s}^{-1} \, \text{cm}^{-2} \, \left( \frac{\dot{M}_{\text{acc}}}{10^{-8}M_\odot \, \text{yr}^{-1}} \right) \\ 
& \times \left( \frac{M_\star}{0.5 M_\odot} \right) \left( \frac{R_\star}{2R_\odot} \right)^{-3} \left( \frac{f}{0.01} \right)^{-1}
\end{aligned}
\end{equation}
with $R_\star$ = 2.48\,$R_{\odot}$ and $M_\star$ = 0.7\,$M_{\odot}$, we calculate the corresponding mass accretion rates for each observation. They range from log\,($\dot{M}_{\text{acc}}$[$M_\odot$\,yr$^{-1}$]) = -8.78 to log\,($\dot{M}_{\text{acc}}$[$M_\odot$\,yr$^{-1}$]) = -7.04 (see Table\,\ref{TableMdot}). 

\begin{figure}[hbtp]
\centering
\includegraphics[scale=0.55]{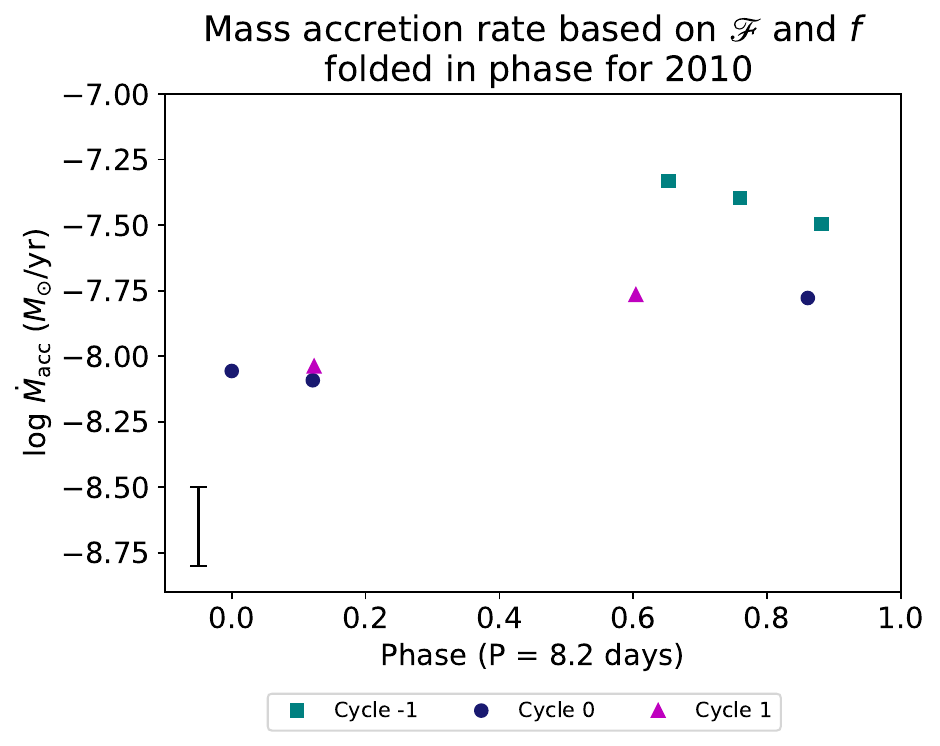}
\includegraphics[scale=0.55]{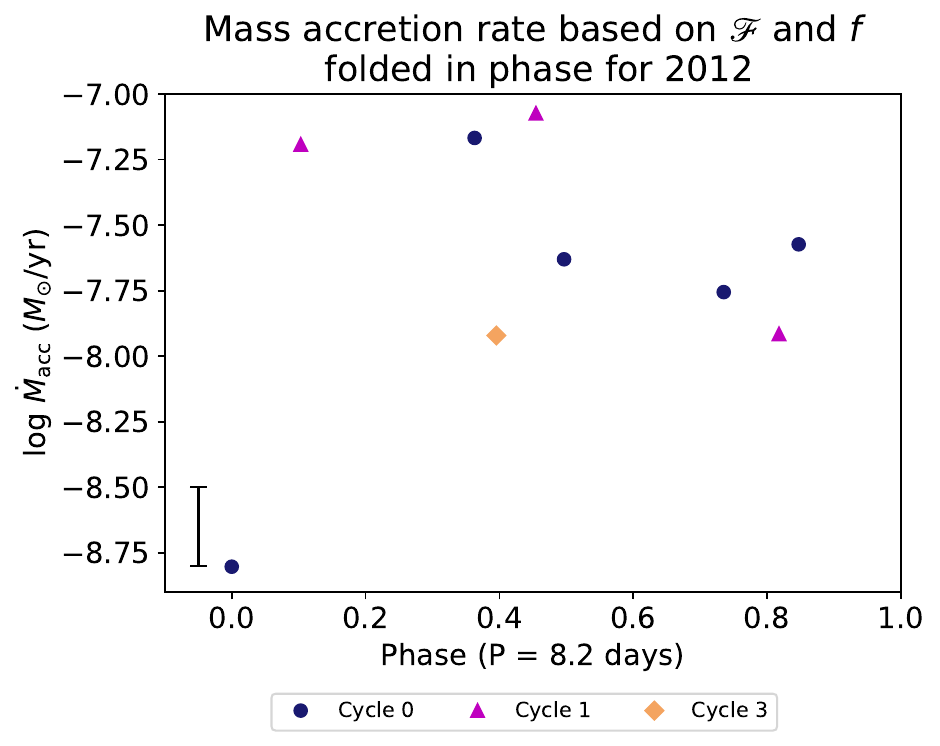}
\caption{Mass accretion rate based on $\mathscr{F}$ and  $f$ over time, shown folded in phase with an 8.2\,day period, for the 2010 (top panel) and 2012 (bottom panel) data. Cycle 0 for the 2010 epoch starts on December 17, while for the 2012 epoch it starts on the first observation of 2012, as in previous figures. Different colors and symbols represent different rotation cycles. The uncertainty on the y-axis is shown on the lower left (see text for details). \label{logMacc_via_F_ff}}
\end{figure}

Figure\,\ref{logMacc_via_F_ff} shows the mass accretion rate obtained via $\mathscr{F}$ and $f$ folded in phase.  We estimated the uncertainty on the values of the mass accretion rate by comparing different models with the distribution of points for the observed veiling. We find uncertainties on the product of the energy fluxes $\mathscr{F}$ and the filling factors $f$, which translates into uncertainties on the mass accretion rate, that range from a factor 2 (or lower for some nights) to a factor 5 (on the nights where the scatter on the observed veiling is the largest). 

Over the range of observations, we see quite a significant range of values for $\mathscr{F}$, the energy flux of the column that impinges on the star, as it varies by two orders of magnitude, because of the range of slopes displayed by the observational veiling. The product of $\mathscr{F}$ and $f$ varies over a smaller range, because the mass accretion rate does not change as much as the energy flux. 

\begin{table*}
\begin{center}
\caption{Accretion luminosities, mass accretion rates based on $L_{\text{acc}}$, energy fluxes, filling factors and mass accretion rates based on $\mathscr{F}$ and $f$ for the 2010 and 2012 epochs. \label{TableMdot}}
\begin{tabular}{c c c c c c}
\hline 
\hline
 Date & log $L_{\text{acc}}$ (err) & log $\dot{M}_{\text{acc}}$ (err) & $\mathscr{F}$ & $f$ & log $\dot{M}_{\text{acc}}$ \\
(yyyy-mm-dd) & ($L_{\odot}$) & ($M_\odot$\,yr$^{-1}$) & (erg\,s$^{-1}$\,cm$^{-2}$) & & ($M_\odot$\,yr$^{-1}$) \\
                       &                       & based on $L_{\text{acc}}$ & & & based on $\mathscr{F}$ and $f$ \\
\hline 
  2010-12-14 & -0.78 (0.16) & -7.62 (0.16) & 7.71 $\times$ $10^{10}$ & 0.046 & -7.29 \\  
  2010-12-15 & -0.84 (0.14) & -7.68 (0.14) & 7.12 $\times$ $10^{10}$ & 0.044 & -7.35 \\  
  2010-12-16 & -0.98 (0.20) & -7.82 (0.20) & 9.45 $\times$ $10^{10}$ & 0.026 & -7.45 \\  
  2010-12-17 & -1.36 (0.20) & -8.20 (0.20) & 2.06 $\times$ $10^{10}$ & 0.033 & -8.01 \\  
  2010-12-18 & -1.33 (0.21) & -8.16 (0.21) & 1.30 $\times$ $10^{10}$ & 0.048 & -8.05 \\  
  2010-12-24 & -1.01 (0.22) & -7.85 (0.22) & 3.63 $\times$ $10^{10}$ & 0.037 & -7.72 \\  
  2010-12-26 & -1.18 (0.16) & -8.02 (0.16) & 1.68 $\times$ $10^{10}$ & 0.043 & -7.98 \\  
  2010-12-30 & -1.11 (0.23) & -7.95 (0.23) & 1.87 $\times$ $10^{10}$ & 0.072 & -7.71 \\  
  \hline
  2012-11-25 & -1.31 (0.27) & -8.15 (0.27) & 1.00 $\times$ $10^{9}$  & 0.117 & -8.78 \\       
  2012-11-28 & -0.82 (0.22) & -7.66 (0.22) & 1.77 $\times$ $10^{11}$ & 0.030 & -7.12 \\     
  2012-11-29 & -0.77 (0.30) & -7.61 (0.30) & 3.04 $\times$ $10^{10}$ & 0.059 & -7.59 \\
  2012-12-01 & -0.92 (0.28) & -7.76 (0.28) & 1.30 $\times$ $10^{10}$ & 0.106 & -7.70 \\
  2012-12-02 & -0.93 (0.31) & -7.77 (0.31) & 3.63 $\times$ $10^{10}$ & 0.059 & -7.51 \\
  2012-12-04 & -0.74 (0.35) & -7.57 (0.35) & 7.71 $\times$ $10^{10}$ & 0.065 & -7.14 \\
  2012-12-07 & -0.57 (0.22) & -7.40 (0.22) & 2.15 $\times$ $10^{11}$ & 0.030 & -7.04 \\
  2012-12-10 & -1.24 (0.20) & -8.07 (0.20) & 1.49 $\times$ $10^{10}$ & 0.063 & -7.87 \\
  2012-12-23 & -0.77 (0.24) & -7.61 (0.24) & 8.50 $\times$ $10^{9}$   & 0.109 & -7.88 \\
\hline
\end{tabular}
\end{center}
\end{table*}


\section{Discussion} \label{sec:discussion}

\subsection{Accretion luminosity and optical veiling} \label{LaccComp}

\begin{figure}[hbtp]
\centering
\includegraphics[scale=0.55]{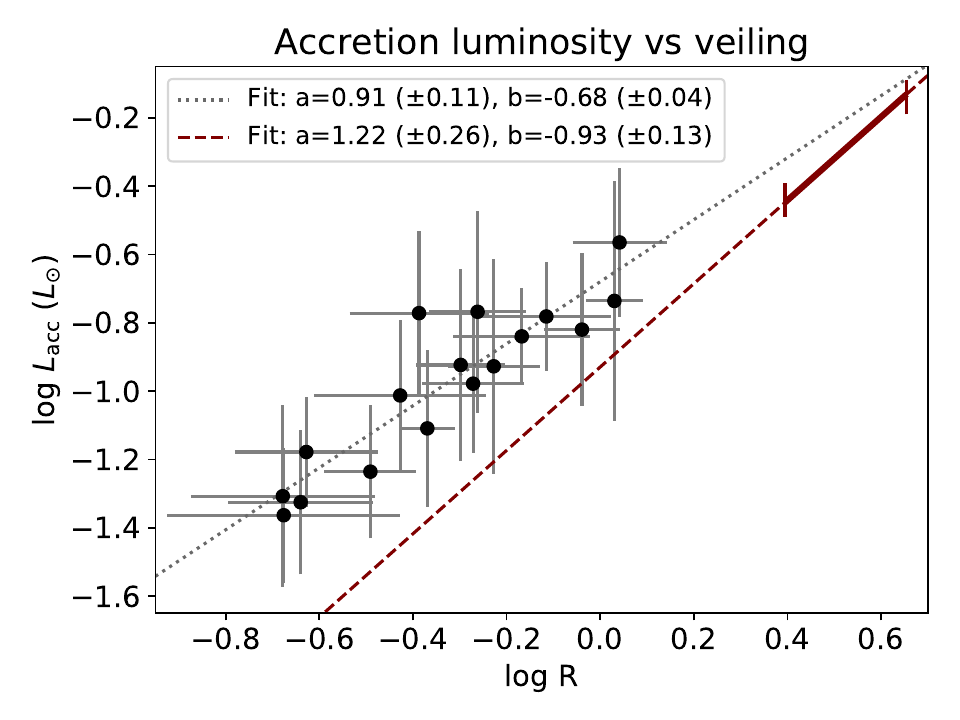}
\caption{Accretion luminosity as a function of the veiling. The black dots show the range of values for DK\,Tau, the gray dotted line is the fit. The maroon dashed line shows the trend found for RU\,Lup by \cite{2022A&A...668A..94S}, with the continuous maroon line highlighting their range. \label{logLacc_vs_logVeiling_Comp}}
\end{figure}

Figure\,\ref{logLacc_vs_logVeiling_Comp} shows the accretion luminosities measured from accretion-powered emission lines as a function of the veiling at 617.50\,nm, for both 2010 and 2012 observations. The correlation is quite good, showing that  the optical veiling traces the accretion luminosity very well. This is notably interesting and useful, as the two quantities are measured in two considerably different ways: the veiling is measured at a single wavelength, while the accretion luminosity is a global quantity in terms of wavelength. Consequently, within the uncertainties, by measuring the veiling (i.e., by simply comparing the depth of observed absorption lines to a photospheric template) at one wavelength (in the optical in this case), one can infer the accretion luminosity which is mostly emitted in the UV range and not at optical wavelengths. 

We find a similar correlation between the accretion luminosity and the veiling as \cite{2022A&A...668A..94S} have found for RU\,Lup, a more luminous ($L$ = 1.31\,$L_{\odot}$) and higher accreting cTTs (log\,($L_{\text{acc}}$[$L_\odot$]) $\sim$ - 0.3) than DK\,Tau. In their Fig.\,8, they show a power-law correlation between the accretion luminosity and the veiling from the Li{\sc i} line, with a slope of 1.22 ($\pm$ 0.26) over a range of log\,$R$ going from 0.40 to 0.65. In comparison, for DK\,Tau and using the optical veiling from the photospheric absorption lines, we find a slope of 0.91 ($\pm$ 0.11) over a range of log $R$ going from -0.70 to 0.10. The slopes  agree  within their error bars. Combined, they cover a large range of accretion luminosities, going from log\,($L_{\text{acc}}$[$L_\odot$]) = -0.10 to log\,($L_{\text{acc}}$[$L_\odot$]) = -1.40. The vertical intercept varies between the two stars, with b = -0.93 ($\pm$ 0.13) for RU\,Lup and b = -0.68 ($\pm$ 0.04) for DK\,Tau. 

The discrepancy between the correlations could stem from the comparison of the veiling from the photospheric absorption lines (for DK\,Tau) with the veiling derived from a single line, the Li{\sc i} line, which suffers from differences in the abundance between RU\,Lup and the adopted template. The authors have applied a correction, but note that it has large uncertainties, which could cause a systematic shift in their numbers. 

It would be very interesting to investigate the relation between the accretion luminosity and the veiling in other stars. It would also be interesting to study the relation for a larger range of veiling values. If a similar relation is found for different stars, this could be a great tool when working on deriving quantities from limited data. 

The correlation shown in Fig.\,\ref{logLacc_vs_logVeiling_Comp} between the optical veiling and $L_{\text{acc}}$ is also potentially very interesting when applied to objects with high and uncertain values of the extinction, as veiling does not depend on it, while any method to measure $L_{\text{acc}}$ does \cite[see also][]{2022A&A...668A..94S}.


\subsection{Mass accretion rates} \label{MaccvsMacc}

\begin{figure}[hbtp]
\centering
\includegraphics[scale=0.55]{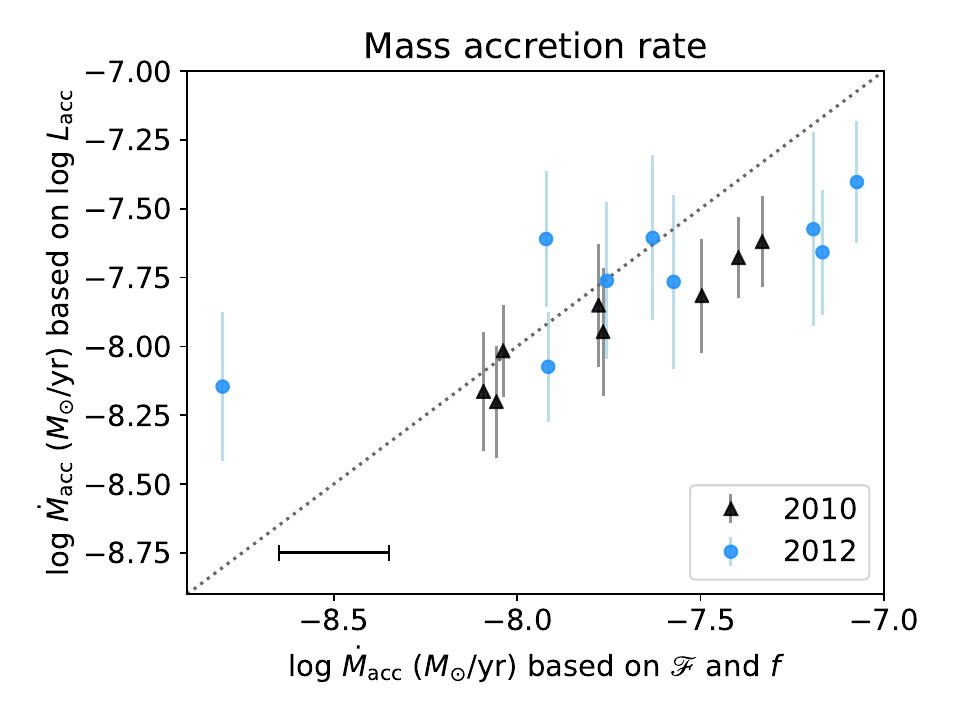}
\caption{Mass accretion rate based on the accretion luminosity as a function of the mass accretion rate based on $\mathscr{F}$ and  $f$, for the 2010 and 2012 data. The gray dotted line shows a one-to-one relationship. The uncertainty on the x-axis is shown on the lower left (see text for details). We note that the accretion shock model method finds almost systematically higher accretion rates than does the accretion luminosity method. \label{logMacc_vs_logMacc}}
\end{figure}

We extracted the mass accretion rate for each night via two different procedures: using the accretion luminosity based on accretion-line powered emission lines (see Sect.\,\ref{Lacc}), and comparing the observed veiling in the optical with accretion shock models to derive values of the energy flux and filling factor (see Sect.\,\ref{ShMod}). The values of $\dot{M}_{\text{acc}}$ that we obtain are listed in Table\,\ref{TableMdot}. They agree within a factor 2 for most nights, which is within the error bars. Figure\,\ref{logMacc_vs_logMacc} plots the two sets of mass accretion rates for both epochs. We can see a trend close to a one-to-one relationship, showing that both methods of estimating $\dot{M}_{\text{acc}}$ are comparable. The discrepancy is probably systematic. Indeed, the mass accretion rate derived using the accretion shock model is almost systematically higher than the one calculated using the accretion luminosity. Similarly, \cite{2022AJ....164..201P} have found systematically higher mass accretion rate from accretion shock models than from the H$\alpha$ luminosity. They mention systematic effects in the modeling methods as the probable cause. 

Shock models have been used to derive accurate values of the mass accretion rates by fitting the observed excess  from the UV to the visual as a combination of shocks of different $\mathscr{F}$ and $f$ \cite[see e.g.,][]{2013ApJ...767..112I, 2021Natur.597...41E}. In the case of DK\,Tau, we find that a single shock model, with $\mathscr{F}$ and $f$ values constrained only by the observed excess emission at optical wavelengths, reproduces quite well the values of $\dot{M}_{\text{acc}}$ obtained from  the intensity of the emission lines. This is possible only when a single shock dominates the observed excess emission at all wavelengths. This is indeed the case not only in DK\,Tau but in 15 of the 21 TTs modeled by \cite{2013ApJ...767..112I}. It is important to note that, as discussed in the following, this result does not rule out models where a single spot has a complex structure \cite[see e.g.,][]{2019ApJ...874..129R, 2021Natur.597...41E}, which can be detected in time-sequence observations. 


\subsection{Variability properties} \label{Varprop}

We investigate the evolution of various quantities with the stellar rotation. Though it should be noted that eight and nine points (for the 2010 and 2012 epoch respectively) are low numbers to detect a trend. The modulation of several of them, such as the accretion luminosity (see Fig.\,\ref{logLacc_Phase}), the mass accretion rate derived from it (see Fig.\,\ref{logMacc_via_Lacc_phase}), the value of the veiling at 617.50\,nm (see Fig.\,\ref{logVeilingTimeFolded}), and the mass accretion rate derived from the shock models (see Fig.\,\ref{logMacc_via_F_ff}) seem to be dominated by the stellar rotation for the 2010 epoch. This is less clear for the 2012 epoch, which experiences some perturbation. \cite{2023A&A...670A.165N} speculate that DK\,Tau may be in the stable accretion regime \cite[see e.g.,][]{2008ApJ...673L.171R, 2013MNRAS.431.2673K} in 2010. This difference between both epochs might be linked to the accretion being less stable in 2012, when the mass accretion rate is higher.  

We observe that the accretion luminosity and energy flux seem to be changing in phase for the 2010 epoch. Consequently, the filling factor has to adjust to the energy flux and the mass accretion rate, as we have demonstrated that both methods of calculating the mass accretion rate (via the accretion luminosity or via the shock models) yield similar results that are changing in phase. The apparent variability of the filling factor with the phase can be used to estimate the location of the accretion spot (see Sect.\,\ref{AccrSpot}). 


\subsection{Accretion spot} \label{AccrSpot}

\begin{figure*}[hbtp]
\centering
\includegraphics[scale=0.14]{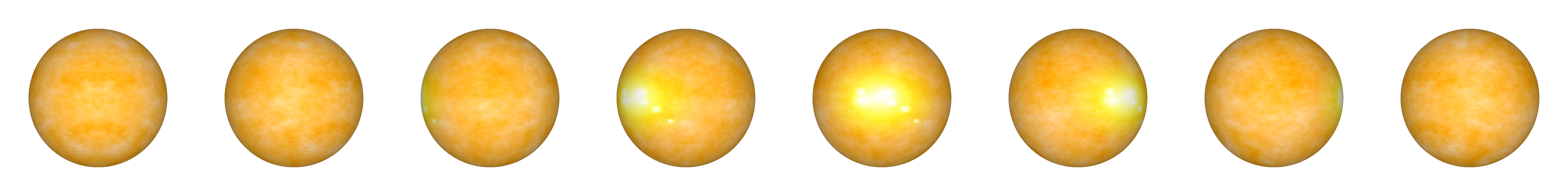}
\caption{Sketch (not to scale) showing an accretion spot at the equator of a star (seen edge on) that is rotating. \label{CDPlot1}}
\end{figure*}

Under the assumption of a single, stable accretion spot, Fig.\,\ref{CDPlot1} shows the rotation of a star seen edge on (i.e., with the observer looking at the page, while the stellar rotation axis is vertical and in the plane of the page), with the spot at the equator (for illustrative purposes). We can see that, as the star rotates, the spot is first invisible, then a portion of it becomes visible, next the whole spot is in view, until it gradually disappears again behind the star. 

\begin{figure}[hbtp]
\centering
\includegraphics[scale=0.14]{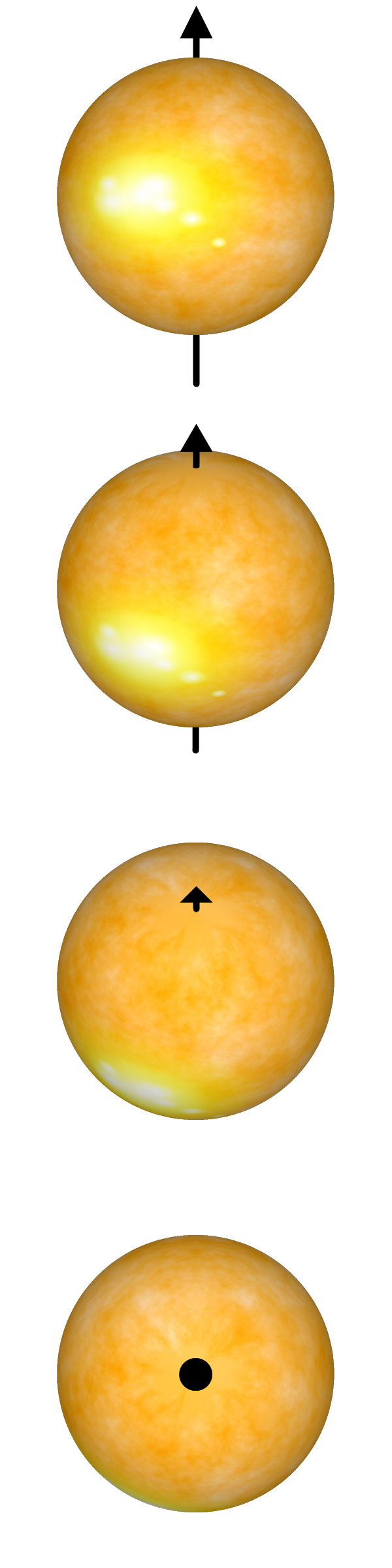}
\caption{Sketch (not to scale) showing a star presenting with an accretion spot and seen with different inclinations (namely 90\si{\degree} or edge-on, 60\si{\degree}, 30\si{\degree} and 0\si{\degree} or pole-one). The arrow represents the rotation axis. The case corresponding to DK\,Tau (i.e., 58\si{\degree}) is closer to the second one. \label{CDPlot2}}
\end{figure}

Figure\,\ref{CDPlot2} shows an accretion spot, at a random location, for different inclinations of the star, with the inclination being the angle of the stellar rotation axis with respect to the line of sight of the observer. The second case is closer to the one corresponding to DK\,Tau's inclination of 58\si{\degree}. We can see that the inclination of the star influences not only the portion of the accretion spot that might be visible, but also affects the perceived area of the spot. This is a geometrical effect, with the true area of the spot remaining unchanged. 

In the case of DK\,Tau, the effects depicted on Fig.\,\ref{CDPlot1} and Fig.\,\ref{CDPlot2} are combined. This means that there are times during the stellar rotation when we do not see the spot, and times when we start seeing it, with a perceived area that is changing due to the inclination of the stellar rotation axis and the rotational phase. 

\begin{figure}[hbtp]
\centering
\includegraphics[scale=0.55]{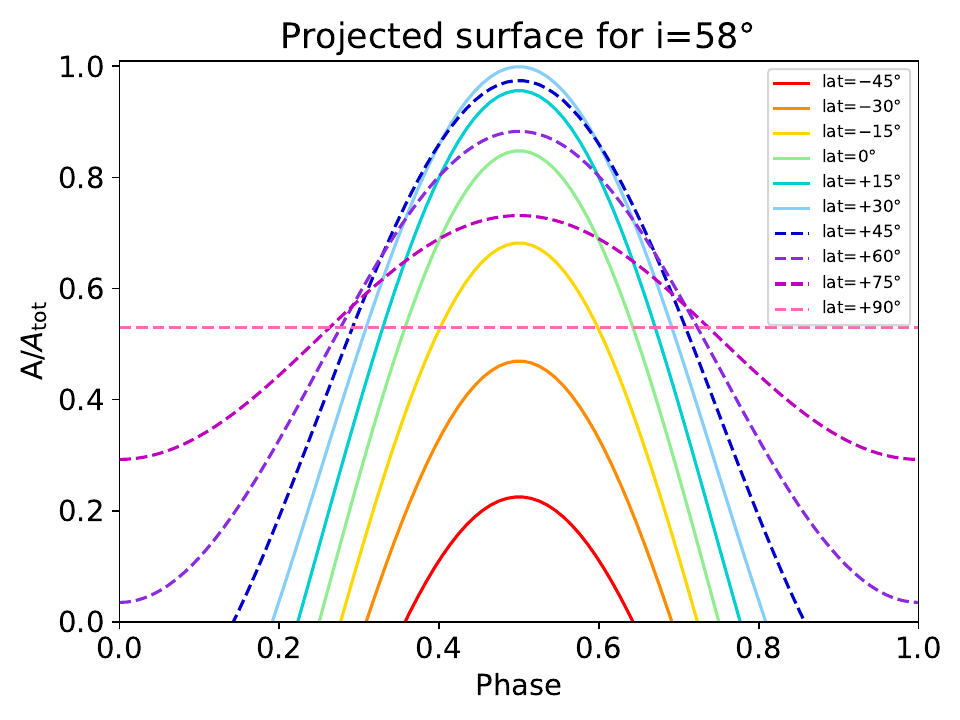}
\caption{Projected surface of an accretion spot as a function of the rotation phase for a star inclined by 58\si{\degree}. Different colors correspond to different locations in latitude for the spot. \label{CDPlot3}}
\end{figure}

Assuming a point-like accretion spot, Fig.\,\ref{CDPlot3} shows the projected surface of the spot (i.e., the perceived area of the spot $A$ divided by the true area of the spot $A_\text{tot}$) at different latitudes, for a star with an inclination of 58\si{\degree}, as a function of the stellar rotation phase. When $A/A_\text{tot}$ = 1, we are seeing the spot in full. Approximating the accretion spot as an ideal lambertian surface, the light it emits in the direction of the line of sight obeys Lambert's law: 
\begin{equation}
I = I_\text{tot} \, \frac{A}{A_\text{tot}} = I_\text{tot} \, \text{cos}\,\alpha
\end{equation}
where $I$ and $I_\text{tot}$ are respectively the perceived and the true luminous intensity of the bright accretion spot,  and $\alpha$ is the angle between the center of the visible stellar disk and the accretion spot. It depends on the inclination $i$ of the star, on the latitude $l$ of the spot and on the stellar rotation phase $p$ via 
\begin{equation}
\text{sin}\,\alpha = \text{sin}\,i \, \text{sin}\,l + \text{cos}\,i \, \text{cos}\,l \, \text{cos}\,p. 
\end{equation}
Because this range of latitudes holds under the assumption of a point-like spot, an extended spot would have the effect of enlarging the wings of the curves, especially the sharp ones. 

By comparing by eye the curves in Fig.\,\ref{CDPlot3} with the shape of the curve for the filling factors (i.e., the fraction of the stellar surface covered by accretion spots) folded in phase (see Fig.\,\ref{ff_Phase}), and assuming that the observed variation of the filling factors are caused by the stellar rotation, we can estimate the location of the accretion spot. For this, we are assuming that we are observing a single, stable accretion spot. In addition, the assumption is that the variation in filling factors depends only on the perceived change of area due to rotation (i.e., that it is only due to a geometrical effect, and is not due to the area of the spot truly changing). We infer that the accretion spot would be located between +45\si{\degree} and +75\si{\degree} in latitude for the 2010 epoch. While the hypotheses made are a simplification of a likely more complex configuration, the curves in Fig.\,\ref{CDPlot3} do seem to reproduce the curve in Fig.\,\ref{ff_Phase}. For the 2012 epoch, however, the perturbation to the rotational modulation prevents a comparison between Fig.\,\ref{CDPlot3} and Fig.\,\ref{ff_Phase}. 

Given our current understanding of magnetospheric accretion, because the magnetic field lines channel the circumstellar material into accretion columns, the magnetic pole is thought to be found at a similar location as the accretion spot, as this is where the matter would accrete. For instance, in the magnetohydrodynamic simulations done by \cite{2021Natur.597...41E} for GM\,Aur, the accretion spot is a few degrees away from the magnetic pole. If we assume that the accretion spot is close to the magnetic pole, we can infer the approximate location of the magnetic pole. For the 2010 epoch, it would be between +45\si{\degree} and +75\si{\degree} in latitude. This translate into a magnetic obliquity (i.e., the orientation of the magnetic field axis compared to the stellar rotation axis) ranging from 15\si{\degree} to 45\si{\degree}. Although this estimation rests on several hypotheses, including the assumption that the observed variation of the filling factors are caused by the stellar rotation, it can be noticed that this estimated range is consistent with the one of 16\si{\degree} to 26\si{\degree} estimated by \cite{2023A&A...670A.165N} for the same epoch and based on measurements of the average line of sight magnetic field. It is also consistent with the magnetic obliquity of 18\si{\degree} for 2011 measured by \cite{2020MNRAS.497.2142M} and based on the radial velocity variability of the He {\sc i} line. In comparison, typical obliquities ranging from 5\si{\degree} to 60\si{\degree} have been measured for other TTs \cite[see e.g.,][]{2014MNRAS.437.3202J, 2018A&A...620A.195A, 2020MNRAS.497.2142M, 2020A&A...642A..99P, 2020MNRAS.491.5660D, 2023MNRAS.520.3049F}. 

Regarding the structure of the accretion spots, the current consensus is that a lower-density area encircles a denser center \cite[see][]{2021Natur.597...41E}. However, a combination of at least two energy fluxes and their filling factors, for a single observation, are required in order to trace two density areas. Since we are fitting each observational veiling with a single shock model defined by one value of the energy flux and filling factor (see Sect.\,\ref{ShMod}), we are approximating the accretion spot structure by a single-density region solely. Extrapolating from previous studies, where the high-density region dominates in the UV range \cite[see e.g.,][]{2013ApJ...767..112I, 2021Natur.597...41E}, we can speculate that with our optical range, we are tracing the low-density part of the spot. This is supported by the fact that the highest energy flux that we find, that is $\mathscr{F}$ = 2.15\,$\times$\,$10^{11}$\,erg\,s$^{-1}$\,cm$^{-2}$, is lower than the value of 1.00\,$\times$\,$10^{12}$\,erg\,s$^{-1}$\,cm$^{-2}$ used in the literature for the high-density region \cite[see e.g.,][]{2021Natur.597...41E, 2022AJ....164..201P}. We may therefore be underestimating the total energy flux by not taking the UV emission into account. 


\section{Conclusions} \label{sec:conclusions}

In this paper, we investigated the accretion onto DK\,Tau, a low-mass classical T\,Tauri star (cTTs), using spectra collected in 2010 and 2012. We studied the veiling (defined as the ratio between the flux coming from the accretion shock and the photospheric flux) across the optical range and found that the peak values (at $\sim$550\,nm) range from 0.2 to 0.9 in 2010, and from 0.2 to 1.3 in 2012. On the nights when the peak values are higher, the slope of the values across the wavelength range is steeper. 

We first calculated the mass accretion rates ($\dot{M}_{\text{acc}}$) for each observation using accretion-powered emission lines and computing the accretion luminosities. We find values ranging from log\,($\dot{M}_{\text{acc}}$[$M_\odot$\,yr$^{-1}$]) = -8.20 to log\,($\dot{M}_{\text{acc}}$[$M_\odot$\,yr$^{-1}$]) = -7.62 in 2010, and from log\,($\dot{M}_{\text{acc}}$[$M_\odot$\,yr$^{-1}$]) = -8.15 to log\,($\dot{M}_{\text{acc}}$[$M_\odot$\,yr$^{-1}$]) = -7.40 in 2012. These values agree with those previously observed \cite[see e.g.,][]{1998ApJ...492..323G, 2011ApJ...730...73F, 2013ApJ...767..112I, 2018ApJ...868...28F, 2022A&A...667A.124G}. 

Additionally, we find, as \cite{2022A&A...668A..94S} have found for the cTTs RU\,Lup, a power-law correlation between the accretion luminosity and the optical veiling. This means that, for DK\,Tau and within the uncertainties, by measuring the veiling at a single wavelength, one can infer the accretion luminosity, which is a global quantity in terms of wavelength. If a similar correlation is found for additional stars, it could be a helpful means of verifying the estimation of the accretion luminosity. Because the veiling does not depend on extinction, it could be particularly useful for high extinction regions. 

We also derived the values of the mass accretion rates by fitting the optical veiling with accretion shock models. These models are characterized by an energy flux ($\mathscr{F}$) which is carried by the accretion column into the shock and is correlated with the density of the material in the column, and a filling factor ($f$) that represents the area of the star covered by the accretion spot. We find that $\mathscr{F}$ ranges from 1.30\,$\times$\,$10^{10}$\,erg\,s$^{-1}$\,cm$^{-2}$ to 9.45\,$\times$\,$10^{10}$\,erg\,s$^{-1}$\,cm$^{-2}$ in 2010, and from 1.00\,$\times$\,$10^{9}$\,erg\,s$^{-1}$\,cm$^{-2}$ to 2.15\,$\times$\,$10^{11}$\,erg\,s$^{-1}$\,cm$^{-2}$ in 2012; while $f$ ranges accordingly from 4.8\,\% to 2.6\,\% in 2010, and from 11.7\,\% to 3.0\,\% in 2012. These values are comparable to the ones mentioned in the literature \cite[see e.g.,][]{2019ApJ...874..129R, 2021Natur.597...41E, 2022AJ....164..201P}. For each night, we are using a single energy flux with its assigned filling factor, making the approximation that one homogeneous accretion spot is dominating the optical emission. The values of $\dot{M}_{\text{acc}}$ extracted from both methods agree within a factor of 2 for most nights. 

Several quantities, including the values of veiling at 617.50\,nm, the accretion luminosity, and the mass accretion rate seem to be rotationally modulated. This is more apparent for the 2010 dataset. The accretion might be more intrinsically variable in the 2012 epoch. For the 2010 epoch, we compared the values of the filling factors folded in phase with curves of the projected surface of a theoretical single, stable accretion spot for a star with DK\,Tau's inclination. We thus estimate that the spot could be located between +45\si{\degree} and +75\si{\degree} in latitude, under the assumption that the observed variation of the filling factors are caused by the stellar rotation. Assuming that the position of the accretion spot is found at a similar location as with the magnetic pole, we infer a magnetic obliquity (i.e., the angle between the magnetic field axis and the rotation axis of the star) ranging from 15\si{\degree} to 45\si{\degree}. This is consistent with the ones estimated by \cite{2023A&A...670A.165N} and \cite{2020MNRAS.497.2142M}. 


\begin{acknowledgements}

The authors thank N. Calvet for helpful discussions and E. Gangi for providing the flux-calibrated spectra of DK\,Tau. 
Based on observations obtained at the Canada-France-Hawaii Telescope (CFHT) which is operated by the National Research Council of Canada, the Institut National des Sciences de l'Univers of the Centre National de la Recherche Scientifique of France, and the University of Hawaii. This project has received funding from the European Research Council (ERC) under the European Union’s Horizon 2020 research and innovation programme under grant agreement No 743029 (EASY: Ejection Accretion Structures in YSOs).

\end{acknowledgements}


\bibliographystyle{aa}
\bibliography{references}


\begin{appendix}

\onecolumn

\section{Optical veiling} \label{VeilingAppendix}

Table\,\ref{TableVeiling} lists the coefficients of the best linear fits ($y = ax + b$) to the observed veiling values.

\begin{table}[H]
\begin{center}
\caption{Fits for the 2010 and 2012 veiling. \label{TableVeiling}}
\begin{tabular}{c c c}
\hline 
\hline
Date & $a$ (err) & $b$ (err)\\
(yyyy-mm-dd) & & \\
\hline 
2010-12-14 & -1.69 (0.45) & 4.61 (1.28) \\       
2010-12-15 & -1.67 (0.45) & 4.48 (1.27) \\     
2010-12-16 & -1.83 (0.37) & 4.83 (1.03) \\
2010-12-17 & -1.05 (0.71) & 2.24 (2.01) \\
2010-12-18 & -0.85 (0.51) & 1.73 (1.45) \\
2010-12-24 & -1.35 (0.63) & 3.33 (1.79) \\
2010-12-26 & -0.95 (0.54) & 2.03 (1.52) \\
2010-12-30 & -1.00 (0.21) & 2.41 (0.61) \\
\hline
2012-11-25 & -0.15 (0.70) & -0.25 (1.97) \\       
2012-11-28 & -2.09 (0.27) & 5.80 (0.76) \\     
2012-11-29 & -1.29 (0.36) & 3.34 (1.01) \\
2012-12-01 & -0.85 (0.36) & 2.07 (1.01) \\
2012-12-02 & -1.36 (0.34) & 3.56 (0.97) \\
2012-12-04 & -1.68 (0.21) & 4.73 (0.58) \\
2012-12-07 & -2.19 (0.36) & 6.15 (1.00) \\
2012-12-10 & -0.90 (0.30) & 2.03 (0.84) \\
2012-12-23 & -0.71 (0.50) & 1.60 (1.41) \\
\hline 
\end{tabular}
\end{center}
\end{table}

\newpage


\section{Accretion luminosity} \label{LaccAppendix}

\begin{figure*}[hbtp]
\centering
\includegraphics[scale=0.39]{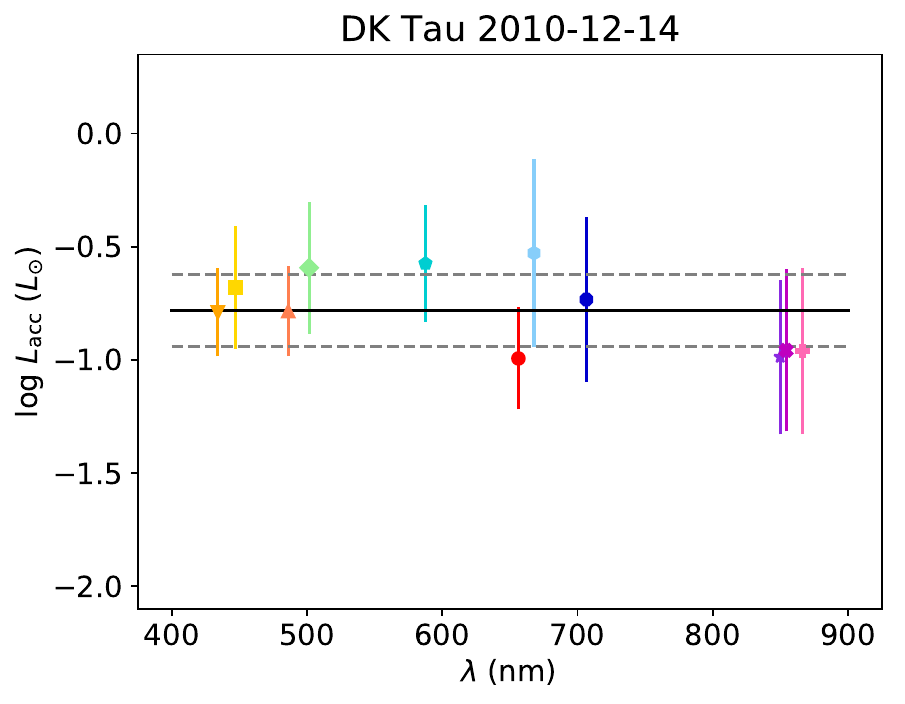}
\includegraphics[scale=0.39]{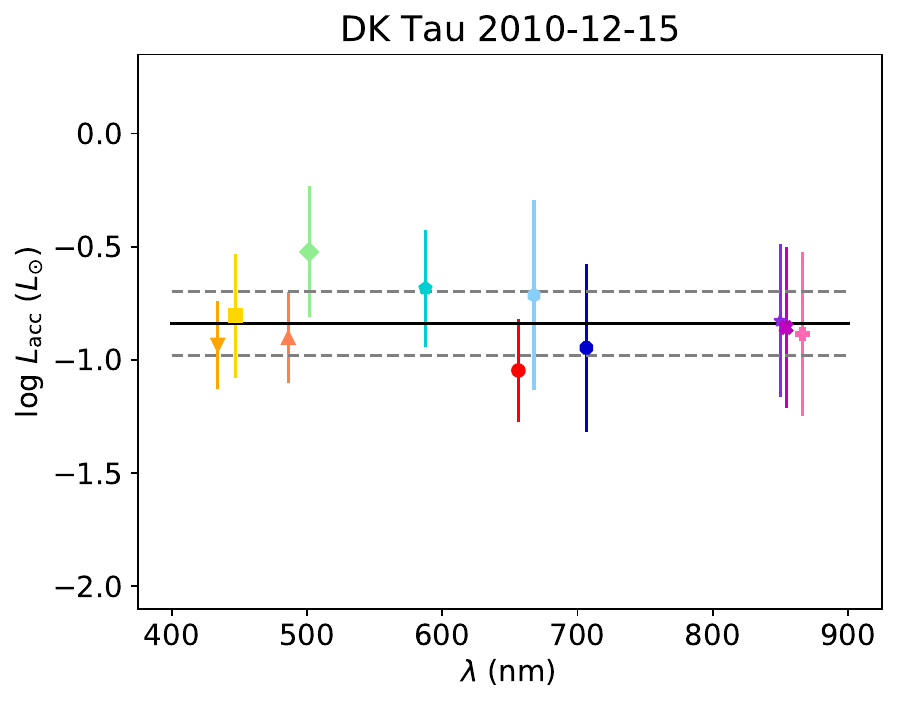}
\includegraphics[scale=0.39]{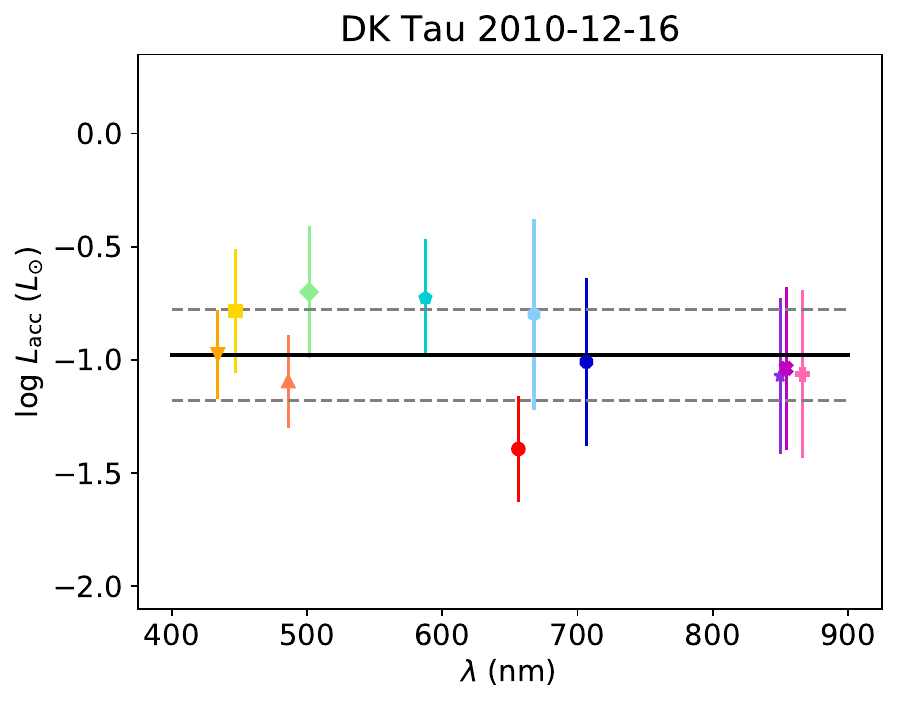}
\includegraphics[scale=0.39]{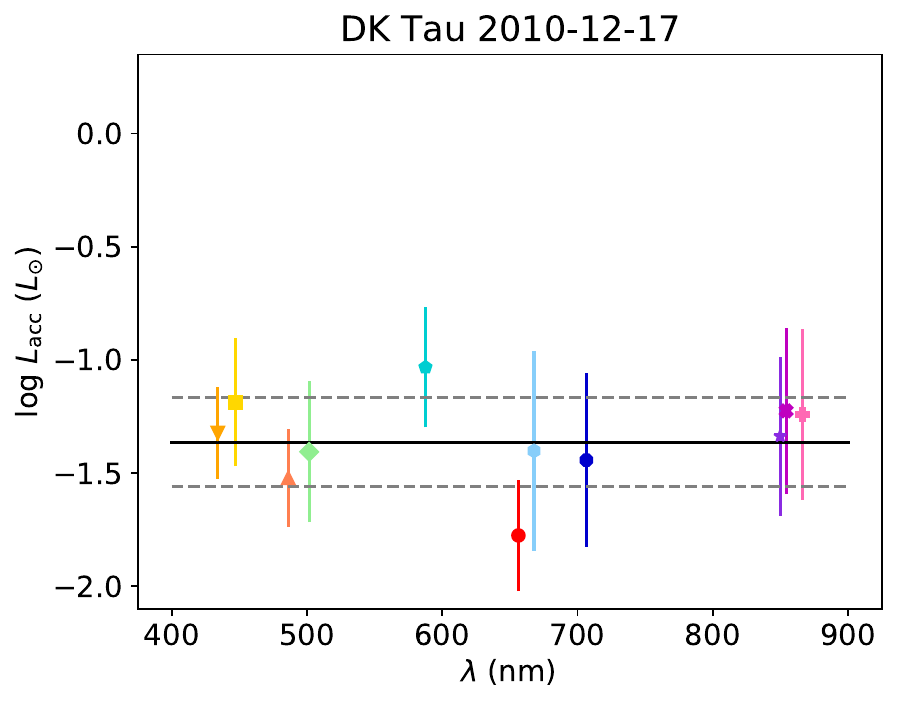}
\includegraphics[scale=0.39]{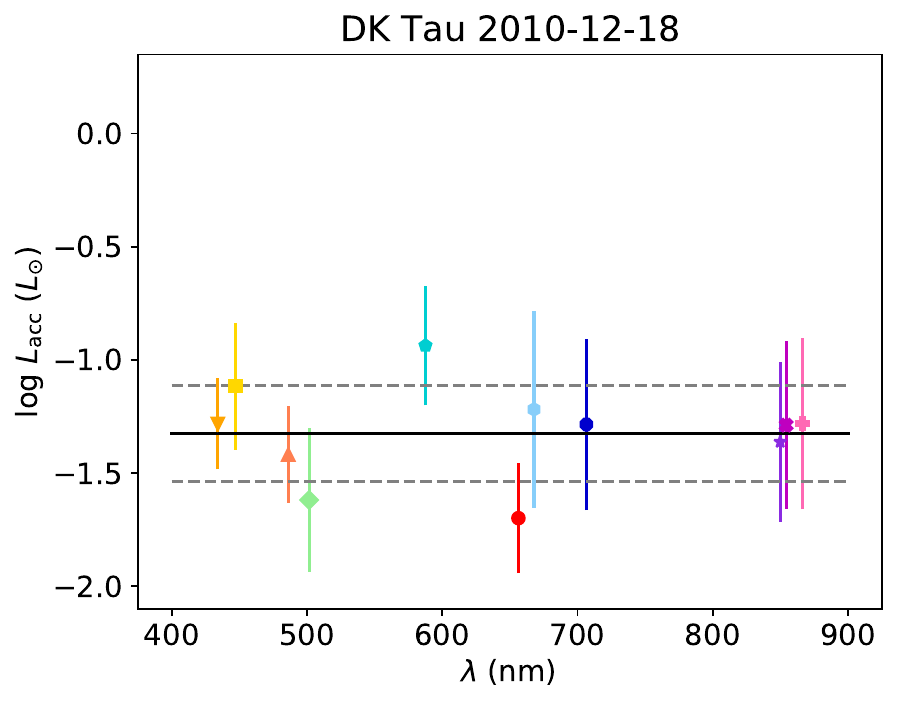}
\includegraphics[scale=0.39]{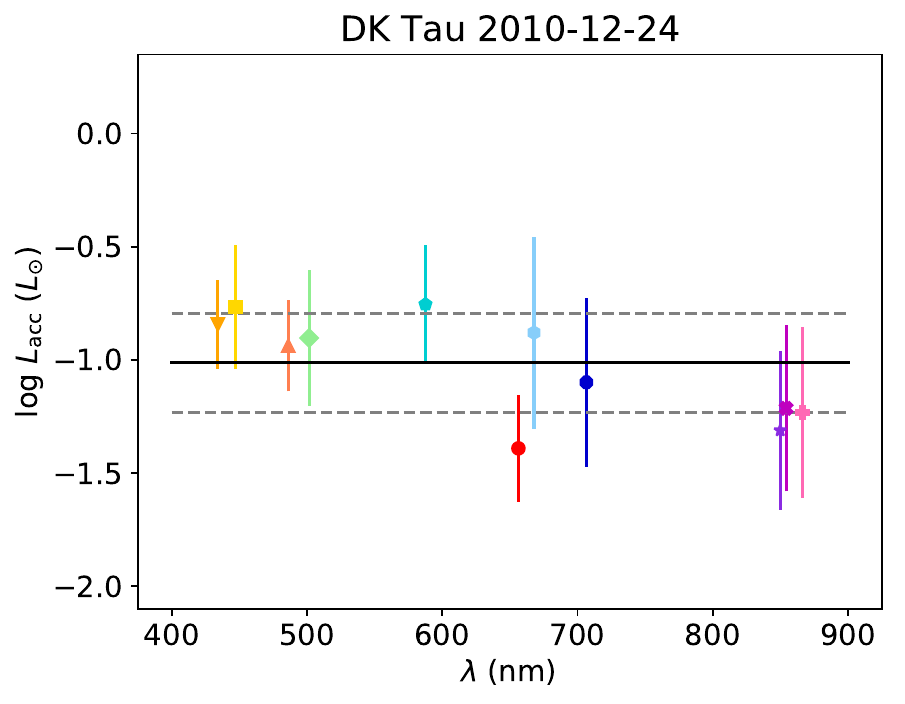}
\includegraphics[scale=0.39]{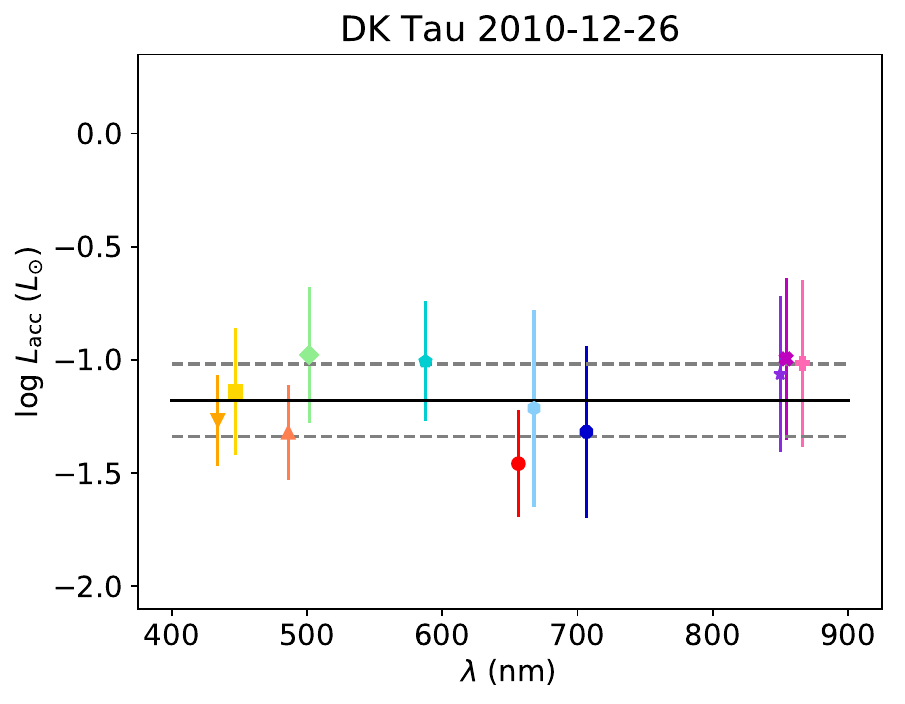}
\includegraphics[scale=0.39]{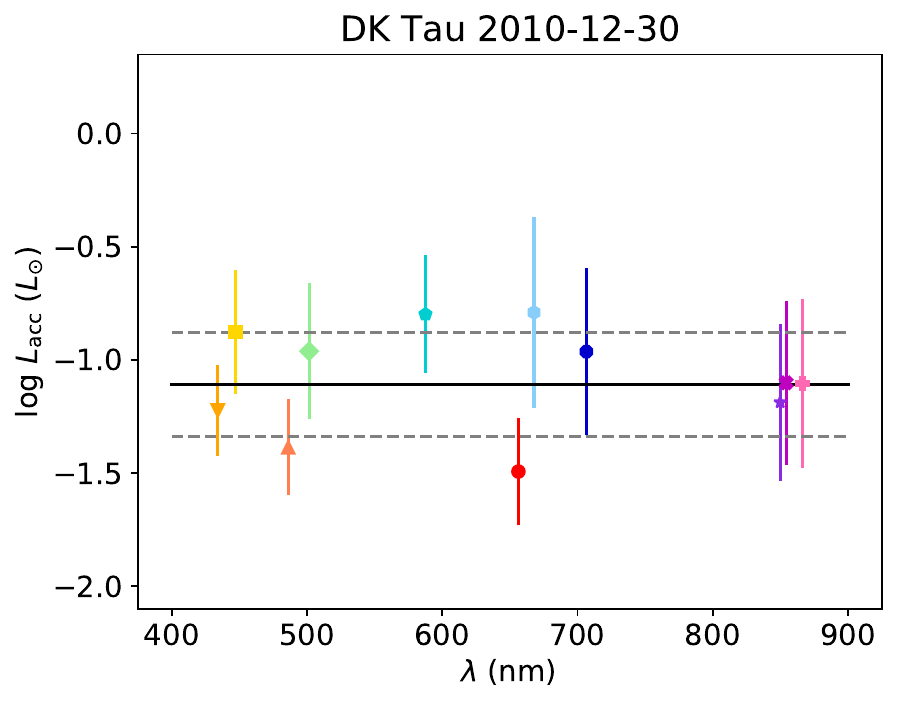}
\includegraphics[scale=0.22]{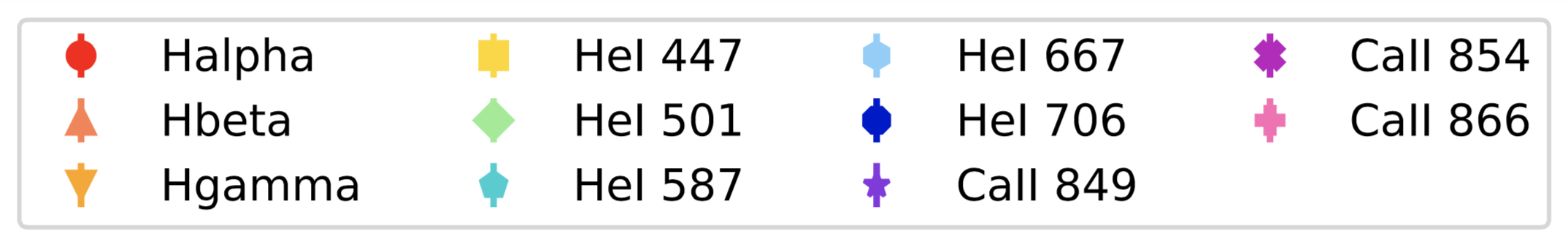}
\caption{Accretion luminosity (colored points), the weighted average (continuous black line) and the spread of values (gray dotted line) as a function of wavelength for the 2010 observations. \label{LaccNight2010}}
\end{figure*}

\begin{figure*}[hbtp]
\centering
\includegraphics[scale=0.39]{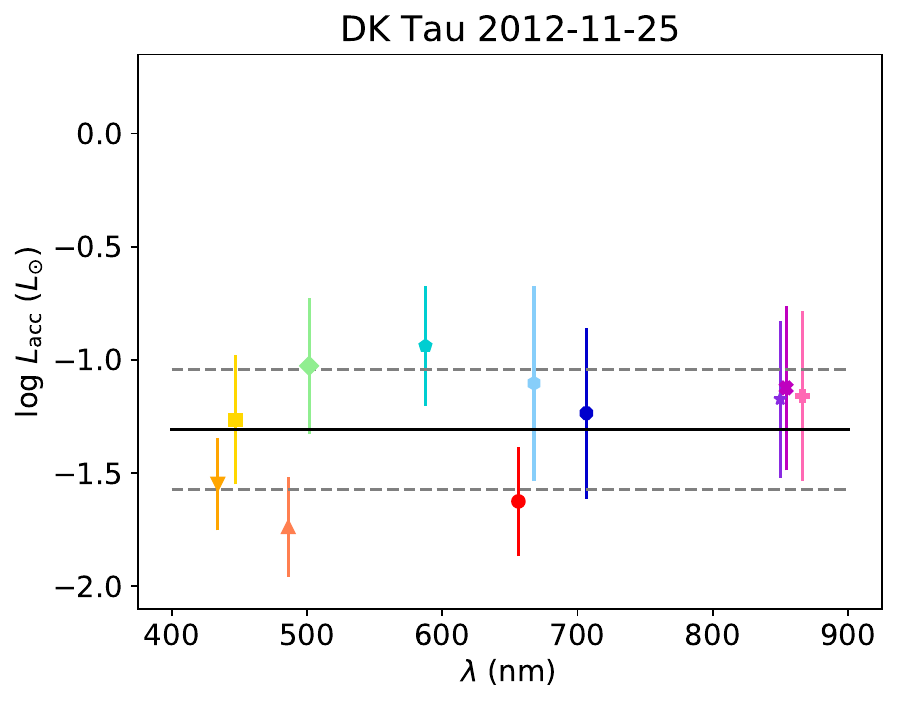}
\includegraphics[scale=0.39]{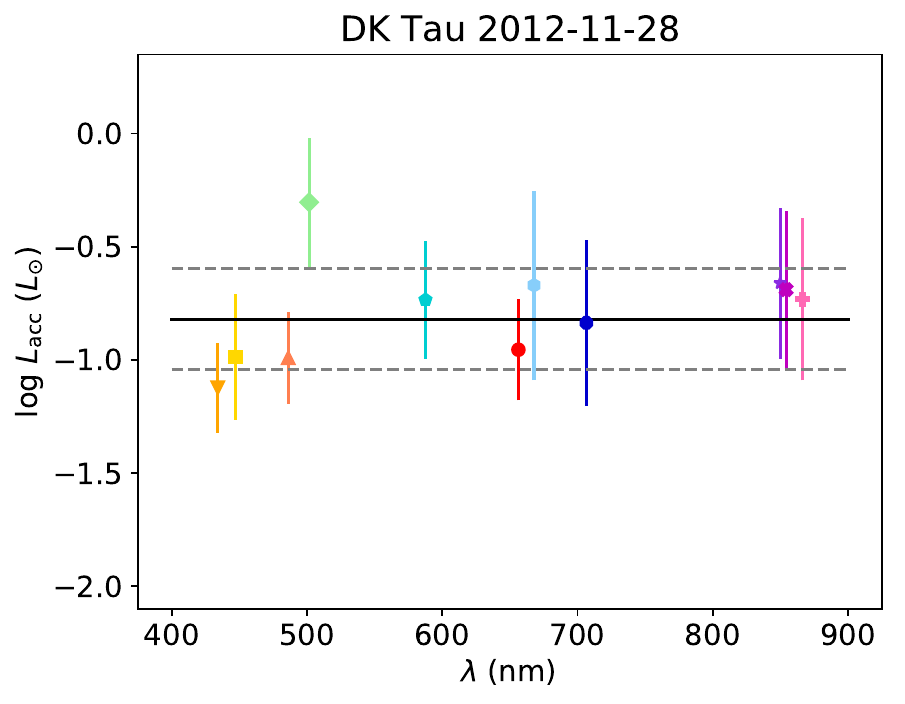}
\includegraphics[scale=0.39]{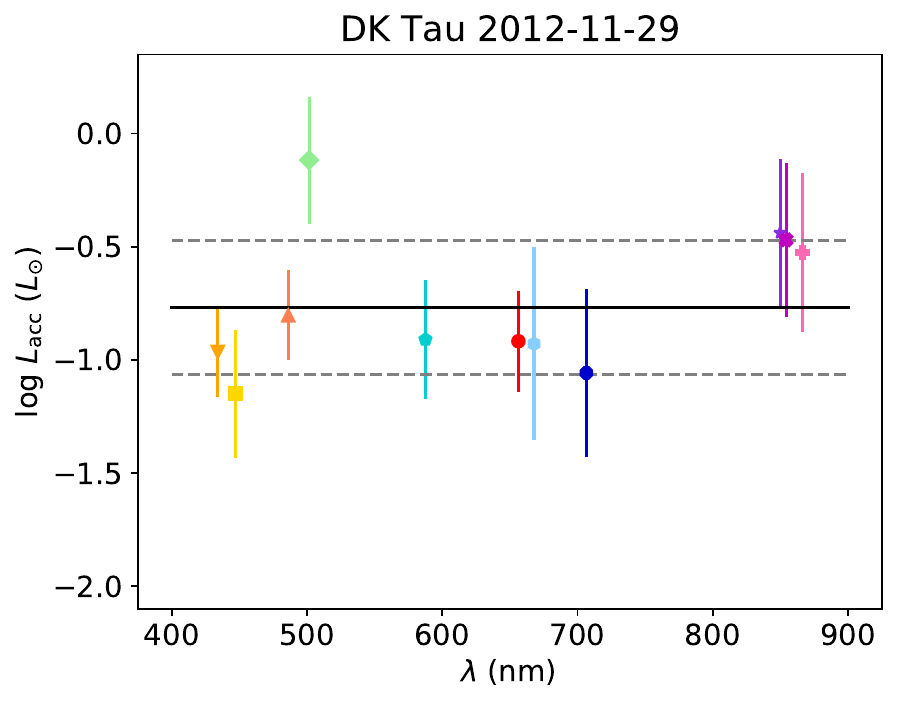}
\includegraphics[scale=0.39]{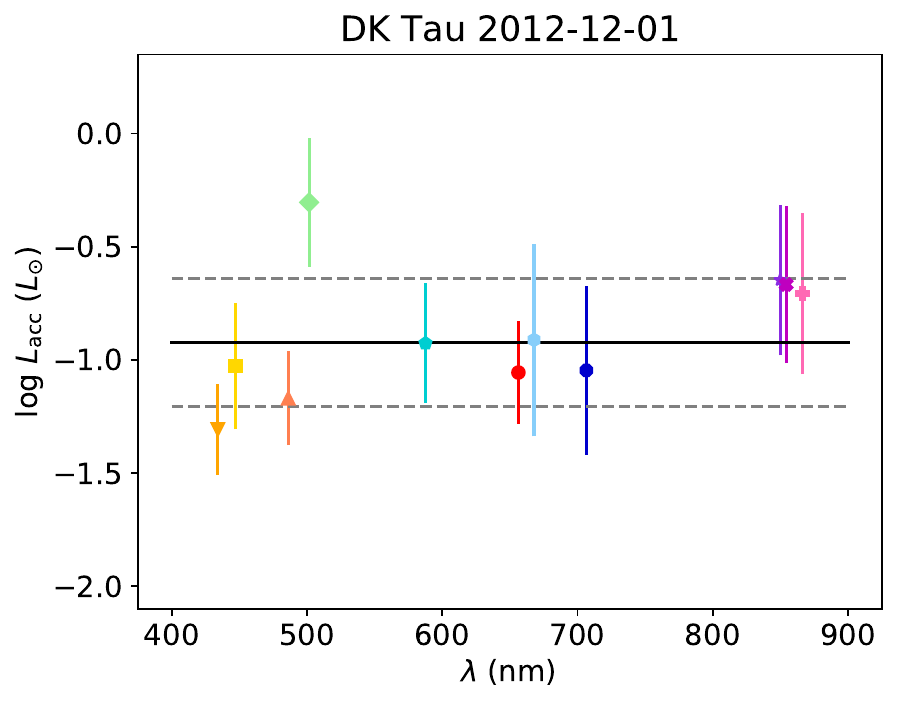}
\includegraphics[scale=0.39]{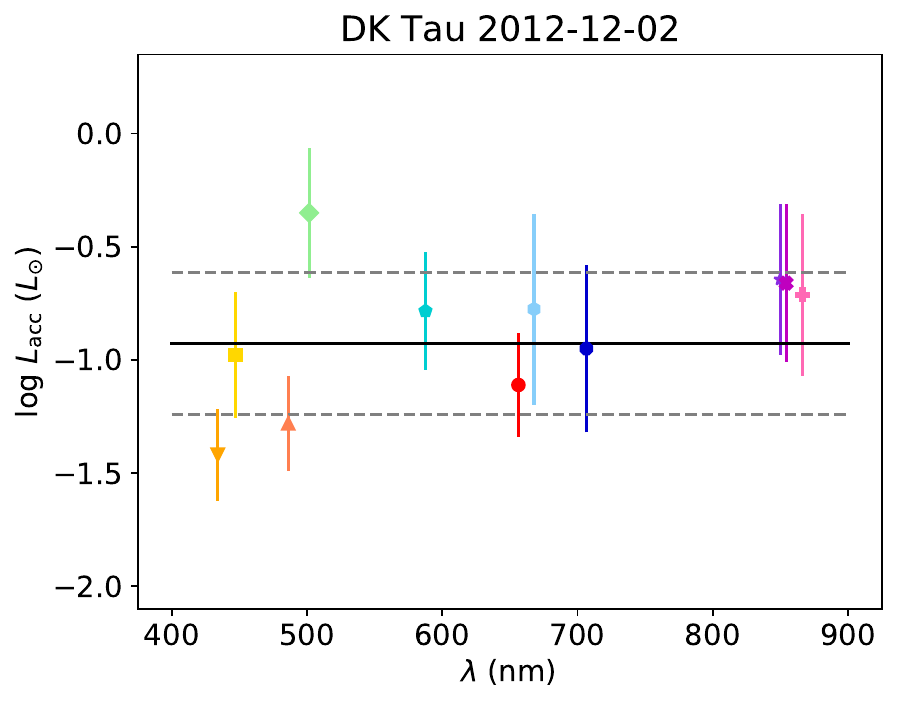}
\includegraphics[scale=0.39]{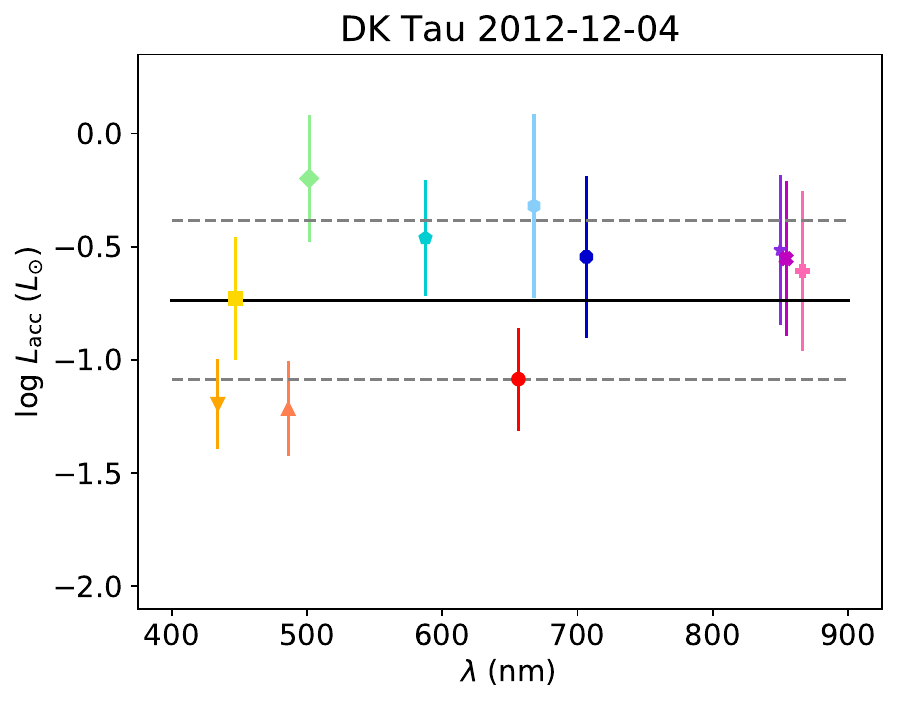}
\includegraphics[scale=0.39]{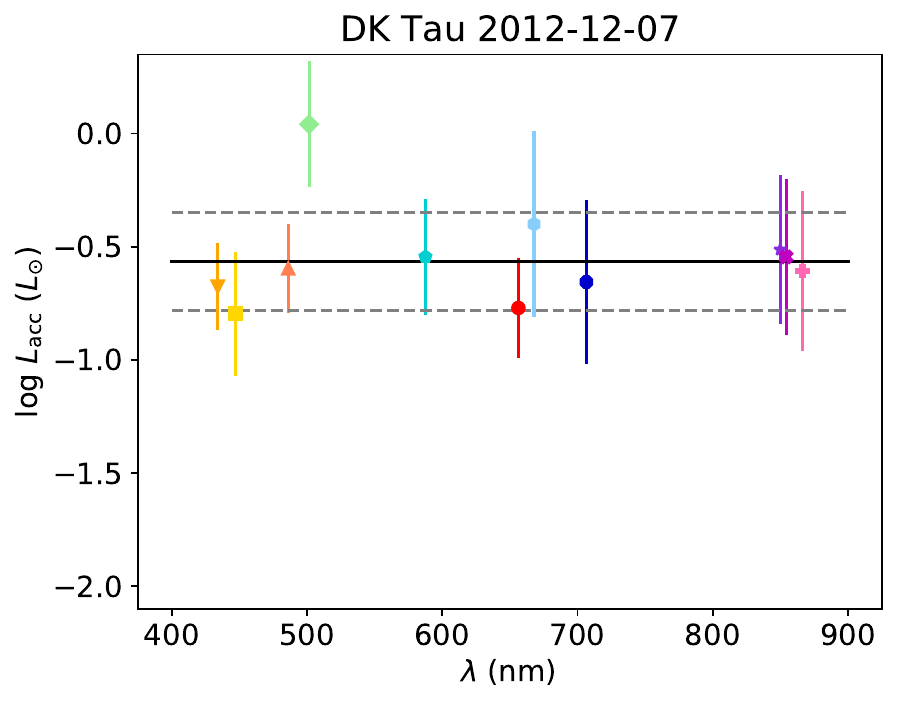}
\includegraphics[scale=0.39]{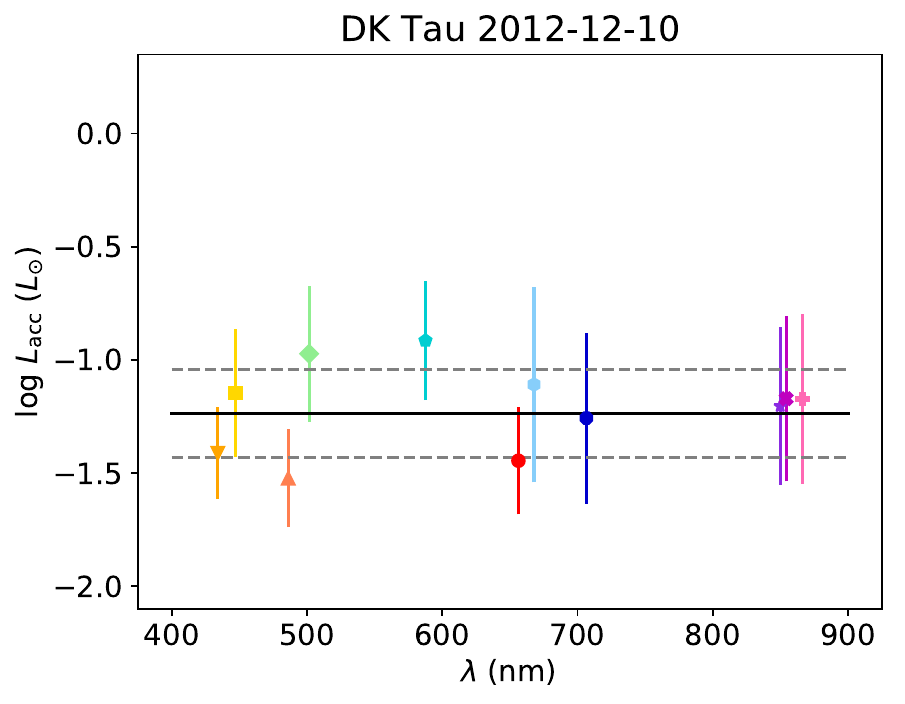}
\includegraphics[scale=0.39]{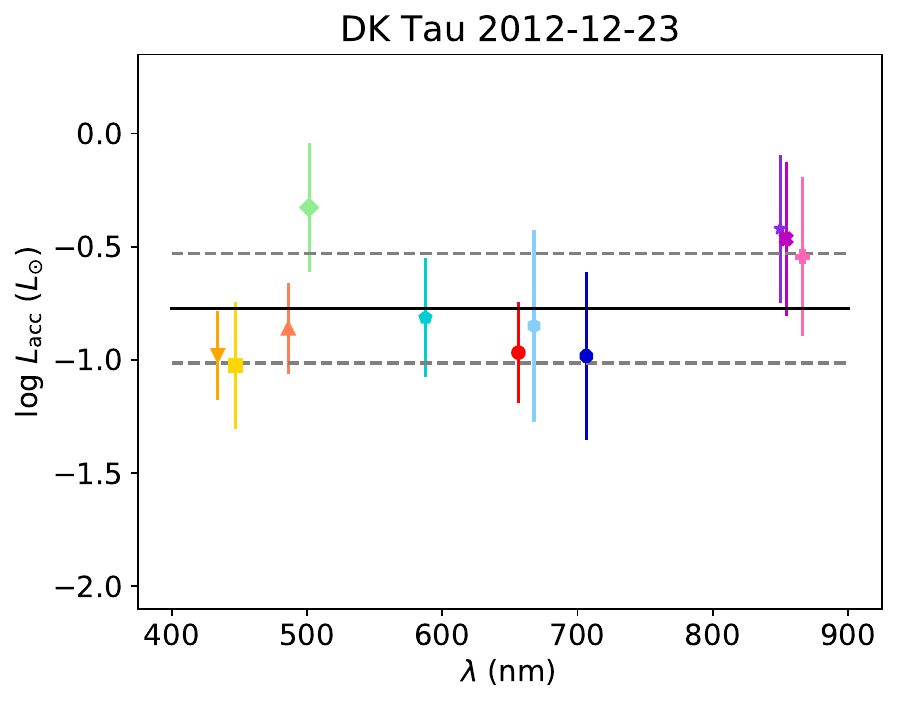}
\includegraphics[scale=0.22]{Figure_Lacc_Night/Screenshot_2023_01_09.png}
\caption{Same as Fig.\,\ref{LaccNight2010}, for the 2012 observations. \label{LaccNight2012}}
\end{figure*}

Fig.\,\ref{LaccNight2010} shows the accretion luminosity as a function of wavelength for each night in 2010. Fig.\,\ref{LaccNight2012} does the same for 2012. The continuous black line shows the weighted average and the gray dotted line shows the spread of values (i.e., the weighted standard deviation of the values from the different lines). 

\FloatBarrier


\section{Shock models} \label{ShModAppendix}

\begin{figure}[hbtp]
\centering
\includegraphics[scale=0.39]{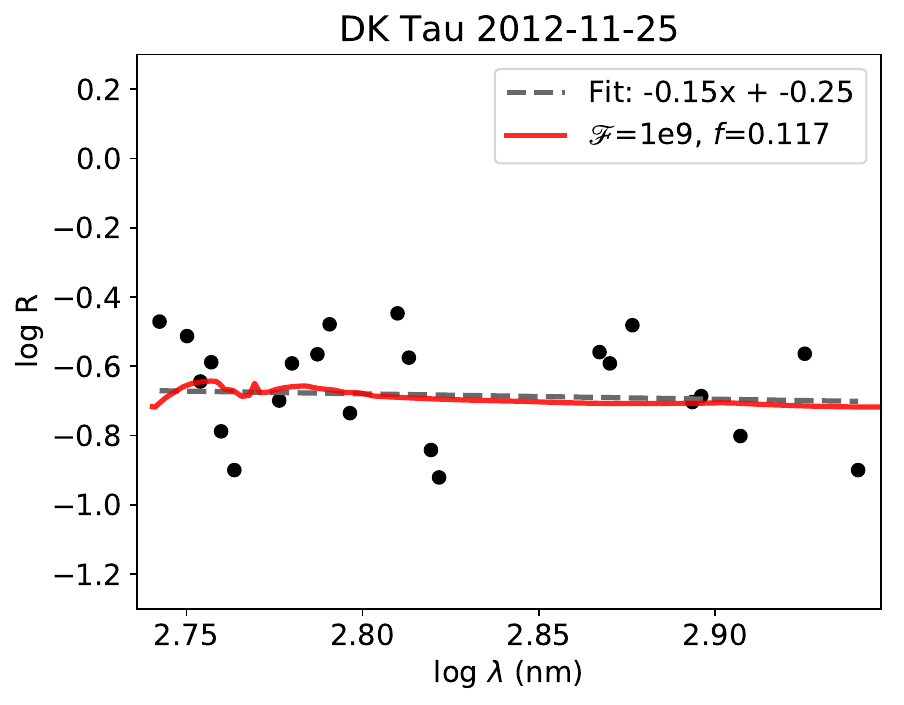}
\includegraphics[scale=0.39]{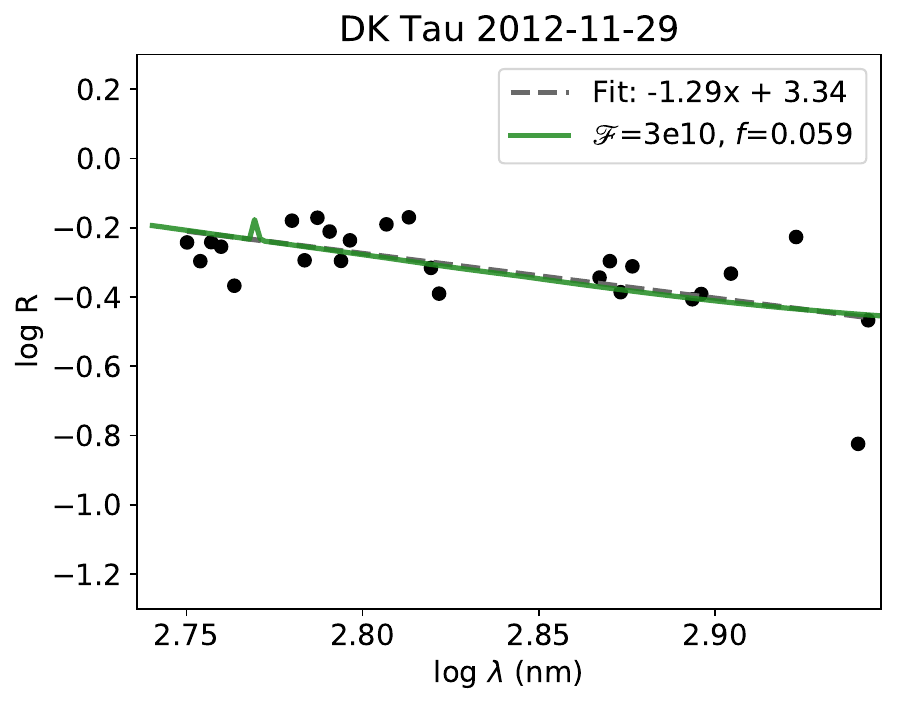}
\caption{Observational veiling (black dots), the best linear fit (gray dashed line) and corresponding modeled veiling (continuous colored line) as a function of wavelength for two examples. \label{Overlap}}
\end{figure}

Fig.\,\ref{Overlap} overlaps the observational veiling (black dots and their best linear fit in gray) with its corresponding modeled veiling (in colors). The modeled veiling is based on an accretion shock model, itself characterized by an energy flux ($\mathscr{F}$) and scaled by a filling factor ($f$). Two examples are given, one with low veiling and the other one with higher veiling. 

\end{appendix}


\end{document}